\documentclass[onecolumn,prl]{revtex4}
\usepackage{amssymb}
\usepackage{amsmath}
\usepackage{color}
\usepackage{graphicx}
\usepackage{subfig}
\def\be{\begin{equation}}
\def\ee{\end{equation}}
\def\bea{\begin{eqnarray}}
\def\eea{\end{eqnarray}}
\usepackage{epstopdf}
\usepackage{epsfig}
\usepackage{multirow}

\begin{document}

\title{Rigorous results of limiting behaviors of total tumor size under cyclic intermittent therapy for the system of reversible phenotype-switchable tumor cells}
\author{Jaewook Joo}
\affiliation{Department of Radiation Oncology, Cleveland Clinic Lerner College of Medicine, Case Western Reserve University, Cleveland, OH, USA}
\begin{abstract}
We are keenly interested in finding the limiting behaviors of total tumor size when tumor cells are subject to the periodic repetition of therapy and rest periods, called intermittent cyclic therapy. We hypothesize that each tumor cell can take either therapy-sensitive or therapy-tolerant phenotype, its phenotype transition is mainly driven by the presence or absence of environmental stress, and such a transition is reversible. Even though those aforementioned hypotheses make the model system simple, most of prior papers attempted to numerically find the optimal therapeutic scheduling that minimizes total tumor size, and there is no rigorous proof of the limiting behaviors of total tumor size to my knowledge. Here we present such long-waited mathematically rigorous results. In the first part of the paper, we present the derivation of total tumor size reduction criterion and prove two theorems of two different limiting behaviors of total tumor size under two different therapy strategies, one leading to an asymptotic finite tumor size according to an iterated map method and anther leading to asymptotically diminishing of total tumor size. In the second part of the paper, we discuss the effects of the intratumoral competition between sensitive and tolerant phenotypes on the total tumor size reduction criterion. 
\end{abstract}
\maketitle
\section{Introduction}

Metastatic cancer is incurable and a leading cause of death, and far less than 10\% of metastatic cancer patients survive five years after initial diagnosis per NIH SEER program. A remarkable emerging new face of this multifaceted disease is its ability to dynamically and reversibly change its phenotype in response to therapy~\cite{weinberg2014}. Upon drug therapy, cancer cells can transition to a drug-tolerant quiescent state and become temporarily resistant to treatment, though its underlying molecular mechanisms are still not clearly elucidated~\cite{shaffer2017,sharma2010,roux2015}. Moreover, cancer therapeutic intervention such as surgery, chemotherapy, and radiation, are all known to accelerate tumor progression, metastasis, and tumor evolution for acquired treatment resistance~\cite{gatenby2020}. This non-static nature of cancer cells presents a huge challenge to current cancer treatment modalities which continue to regard cancer as a static target.  

The aforementioned stress-induced tumor progression and treatment resistance acquirement prompted to re-evaluate the current therapeutic effects on tumor biology and physiology~\cite{gatenby2020}. In an attempt to eliminate tumor, the most aggressive therapeutic modality is oftentimes elected in the clinical settings, yet it pushes tumor to its extreme corner, driving it to be more stem cell-like, whose behaviors resemble those of the aggressive, uncooperative, unregulated, and resilient ancient single cell types~\cite{davies2011}. Understanding of this dynamical response of tumor, namely cell plasticity and phenotype switching, to external insult is critically important for the development of better therapeutic outcomes particularly for metastasis. 

Evolutionary Adaptive Therapy (EAT) is an emerging and promising therapeutic approach which treats cancer as a moving target and as interacting species in tumor-immunity-tissue ecological settings and utilizes the key principles of fully matured fields of ecology and evolution. EAT can be summarized with three key phrases: anticipation of target movement, repetitive therapy with the intent of tumor containment, and minimal necessary dose instead of maximum tolerated dose~\cite{gatenby2020}. It borrows its strategy and ideas from Integrated Pest Management where, given the information on the life cycles of pests and their interaction with the environment, the most economical (thus minimal) pest control methods are implemented to manage pest damage with the least possible hazard to the environment and the least possible pest resistance\cite{gatenby2020}. Intermittent Cyclic Therapy (ICT), the most popular and conventional strategy of EAT, uses the alternating periods of treatment and treatment holidays to defer the timing of the uprising of untreatable resistant cancer phenotype above a clinically acceptable threshold by taking advantage of the intratumoral competition through which the prevailing number of fast-growing Tx-sensitive phenotype suppresses the growth of slowly growing Tx-resistant phenotype. ICT strategy has been successfully applied to radiation therapy for Platelet-derived growth factor-driven glioblastoma cell lines, hormone therapy for metastatic castrate-resistant prostate cancer patients, and chemotherapy for in vitro human breast cell lines and mouse in vivo model~\cite{atlock2015,leder2014,goldman2015,zhang2017,zhang2022}.   

Lots of prior mathematical and clinical studies handled the therapy-induced resistance issues. The primary goals of their studies have been tumor size reduction, tumor progression delay, and tumor size containment. Most of prior mathematical studies dealt with irreversible genetic mutation-induced resistance and relied on the optimal control theory-based numerical simulations to present the optimal treatment scheduling. Those efforts were nicely reviewed and  analyzed in the reference~\cite{noble2021}. The key principles that enable tumor containment are the intratumoral competition between treatment-sensitive and treatment-resistant tumor phenotypes. To my knowledge, there is no mathematically rigorous analysis of the limiting behaviors of tumor size for the system of reversible phenotypic switching tumor cells despite its simplicity and a great amount of insight that one can obtain from this analysis. 

We present the rigorous results of the limiting behaviors of tumor size under the deterministic ICT. We consider an ordinary differential equation model of therapy-induced phenotypic switching between Treatment-sensitive and Treatment-tolerant phenotypes which undergo ICT with alternating treatment and Tx- holidays. We aim to investigate the effects of therapy-induced cancer phenotypic switching on tumor size reduction and present two strategies of ICT. We also discuss the effects of the intratumoral competition between sensitive and tolerant phenotypes on the total tumor size reduction criterion.

\section{Model description for stress-induced phenotypic switching during one cycle of intermittent therapy}
We aim to investigate the effects of stress-induced phenotypic switching of tumor cells on therapeutic efficacy and intend to propose a therapy strategy that yields the maximal tumor size reduction. We consider a simple mathematical model for the growth of a population of well-mixed tumor cells which are capable of reversibly switching between two phenotypic states, sensitive/growing and tolerating/quiescent states, in response to time-dependent change of external stress. 
\begin{eqnarray}
\frac{dS}{dt} &=& \rho_s S -\alpha(\delta) S +\beta(\delta) R- \delta(t) S \\ 
\frac{dR}{dt} &=& \alpha(\delta) S -\beta(\delta) R 
\end{eqnarray}
$S$ represents the density of the sensitive/growing cancer cells that grows at a constant growth rate of $\rho_s$ regardless of environmental stress and are susceptible to therapy-induced killing at the time-dependent rate $\delta(t)$. $R$ represents the density of the tolerating/quiescent cancer cells which don't grow, but can tolerate/survive a therapy-induced environmental stress. The sensitive cells switch to the tolerating cells at a stress-dependent forward switching rate of $\alpha (\delta)$ and the tolerating cells reversely switch back to the sensitive cells at a stress-dependent backward switching rate of $\beta(\delta)$. Both forward and backward switching rates of $\alpha(\delta)$ and $\beta(\delta)$ depend on the level of therapy-induced stress and thus are the functions of the time-dependent therapy-induced killing rate of $\delta(t)$. We assume that the forward switching rate takes the "sigmoidal" shape as a function of therapy-induced stress level whereas the backward switching rate is of the reverse sigmoidal shape. For the simplicity of the model, we make a further assumption that the therapy-induced stress level is much larger than $K_{\alpha}$ and $K_{\beta}$ (defined as the value of stress at which the forward and backward switching rates are the half of their maximum rates) and thus the forward (backward) switching rate can be approximated to be a constant maximum (minimum) value, i.e., $\alpha(\delta) = \alpha$ and $\beta(\delta)=0$ for the duration of therapy. However, during intermission between two consecutive therapy sessions, we assume that the system is immediately brought back to a stress-free environmental condition without any time-delay for its restoration and its stress level is much lower than $K_{\alpha}$ and $K_{\beta}$ and the forward (backward) switching rate is approximated to be a constant mininum (maximum) value, i.e., $\alpha(0)=0$ and $\beta(0)=\beta$ for the duration of stress-free environmental condition. In our simple model, the stress-free and the stressful periods, defined as $T_{f}$ and $T_{s}$, periodically alternate: $T_{f} \rightarrow T_{s} \rightarrow T_{f} \cdot \cdot \cdot$. This effectively mimics the clinical cycles of therapy and rest periods. We also assume that the stress-induced killing rate of $\delta(t)$ is simplified to be a constant, $\delta(t)=\delta$ for $t \in T_{s}$ and $\delta(t)=0$ for $t \in T_{f}$. 

The tolerable cells are assumed to be quiescent and this cellular state of no growth (i.e., cell cycle arrest) confer a therapeutic resistance to tolerable cells. The general case of the non-zero growth rate of tolerable cells will be discussed in the section XII. Also, the stress-dependent forward switching rate is assumed to be independent of the stress level during the stressful period. The general case of the stress level (or therapeutic dosage)-dependent forward switching rate will be discussed in the section XI. 

With the above simplification and assumptions, we write two sets of the ordinary differential equations (ODE), one set for stress-free condition and another for stressful condition, respectively. During the stress-free period of the one cycle therapy, for $t \in (0,t_f)$, the time evolution of the sensitive and tolerant tumor populations is governed by the coupled ODE. 
\begin{eqnarray}
\frac{dS(t)}{dt} &=& \rho_s S + \beta R \\
\frac{d R(t)}{dt} &=& -\beta R
\end{eqnarray}
The solution of the system of the coupled ODE with the initial conditions of $S(0)=R(0)=R_0$ is given as follows:  
\begin{eqnarray}
\label{S_Tf}
S(t) &=& R_0 e^{\rho_s t} +\frac{\beta R_0}{\rho_s+\beta} (e^{\rho_s t} - e^{-\beta t}) \\
\label{R_Tf}
R(t) &=& R_0 e^{- \beta t}
\end{eqnarray}
During the stressful period of the one cycle therapy, for $t \in (t_f, t_f+t_s)$, the coupled ODE is given as follows:
\begin{eqnarray}
\frac{dS(t)}{dt} &=& (\rho_s -\delta -\alpha) S   \\
\frac{d R(t)}{dt} &=& \alpha S
\end{eqnarray}
and the solution of the above ODE is provided with the initial conditions of $S(t_f)$ and $R(t_f)$ as below:
\begin{eqnarray}
\label{S_Ts}
S(t) &=& S(t_f) e^{-\rho_{eff}(t-t_f)} \\ 
\label{R_Ts}
R(t) &=& R(t_f) + \hat{\alpha} S(t_f) (1-e^{-\rho_{eff}(t-t_f)})
\end{eqnarray}
where $\rho_{eff}=\delta+\alpha-\rho_s$ and $\hat{\alpha}=\frac{\alpha}{\rho_{eff}}$.

If $\rho_{eff}<0$, the sensitive population size will indefinitely increase in time during therapy. So, we make a reasonable assumption of the positivity of $\rho_{eff}>0$, saying that therapeutic stress is large enough that the sensive population size decreases exponentially and the tolerating population saturates to $R(t_f)+\hat{\alpha} S(t_f)$ in a long time limit of $t_s$. 

\section{Optimal therapy strategies for the extreme cases of infinitely large switching rates and therapy-induced killing rate}

We are keenly interested in the reduction of the total tumor size under this alternating environment. We present two propositions under two extreme conditions, one with infinitely large switching rates and another with infinitely large tumor killing rate. In both cases, we calculate the total tumor size and present the optimal therapy strategy which minimizes the total tumor size. 

{\bf Proposition 1}: In the limit of infinitely fast forward and backward switching rates of $\alpha \rightarrow \infty$ and $\beta \rightarrow \infty$, the total tumor size of $N(t)=S(t)+R(t)$ grows exponentially only during stress-free periods $t_{f}$ and the evolution of total tumor size can be expressed as $N(t) = N(0) e^{\rho_s M t_{f}}$ where $N(0)$ is the initial total tumor size and $M$ is the number of the therapeutic cycles in time $t$. 

{\bf Proof of Proposition 1:} Imagine a sequence of alternating periods of $t_{f}$ and $t_s$. During the first stress-free period of $t_{f}$, all of tolerating cells immediately switch to sensitive cells and there is only a single phenotype of sensitive cells which grows exponentially: $N(t=t_{f})=N(0) e^{\rho_s t_{f}}$. During the next time period $t_s$ of stressful condition, all of sensitive cells immediately switch to tolerating/quiescent cells and there is only a single phenotype of tolerating/quiescent cells whose population size remains constant: $N(t=t_{f}+t_{s})=N(0) e^{\rho_s t_{f}}$. Thus, after $M$ cycles of alternating stress-free and stressful conditions, the total population size is purely determined by the total length of stress-free periods: $N(t) = N(0) e^{\rho M t_{f}}$. In this case, the optimal therapy strategy is to reduce the stress-free period $t_{f}$ to null, i.e., a perpetual stressful environment. However, the downside of this strategy is that it can never reduce the total tumor size below the initial tumor size of $N(0)=S(0)+R(0)$.  
 
{\bf Proposition 2}: In the limit of infinitely large stress-induced cell death rate of $\delta \rightarrow \infty$, the total tumor size decreases exponentially during stress-free periods: $N(t) = R(0) e^{-\beta M t_{f}}$ where $M$ is the number of the cycles of stress-free and stressful periods in time $t$. 

{\bf Proof of Proposition 2:} Imagine a sequence of alternating periods of stress-free and stressful conditions. Let us assume that $S(0)>0$ and $R(0)>0$. During a stress-free period, a fraction of tolerating cells switch to sensitive cells with a finite backward switching rate $\beta$ and tolerating tumor size decreases exponentially: 
$R(t=t_{f})=R(0) e^{-\beta t_{f}}$. However, during a subsequent stressful period, all of sensitive cells are immediately killed without a chance of converting its fraction to tolerating cells and the tolerating population size remains constant: 
$R(t=t_{f}+t_{s})=R(t=t_{f})=R(0) e^{-\beta t_{f}}$. Thus, after M cycles of alternating stress-free and stressful conditions, the total population size is purely determined by the time evolution of tolerating cells, i.e., its exponential decrease during the total length of stress-free periods of $M t_{f}$: $N(t=M(t_{f}+t_{s}))=R(t=Mt_{f})=R(0) e^{-\beta M t_{f}}$. The optimal therapy strategy is to maximize $M t_{f}$, by increasing either the number of cycles $M$ or the stress-free period $t_{f}$.

\section{Intermittent cyclic therapy for finite and non-zero phenotypic switching rates}
We define a single cycle of therapy as a sequence of a rest/stress-free period followed by a therapeutic/stressful period. For mathematical convenience and possibly practical purpose, we demand that, at the beginning and the ending of the cycle, two phenotypes of tumor are of equal size. The therapeutic efficacy after one cycle is measured by the amount of tumor size reduction between the beginning and the ending of the cycle. As described and emphasized in the main text, the rest/stress-free period in our work plays a crucial role for the re-distribution of the tumor phenotypes which  needs to be fine-tuned for the optimal tumor size reduction. We claim and show that this optimization of tumor size reduction after one cycle of therapy can be generalized to find the optimal tumor size reduction after multiple cycles of therapy. In the following sections, we derive the tumor size reduction criterion and calculate the quantity of tumor size reduction after one cycle of therapy and then provide a mathematical description of the limiting behavior of tumor size after multiple cycles of therapy by using the iterated map. Then, we propose the optimal therapeutic strategy that can eliminate tumor after a large number of therapeutic cycles. Finally, we discuss how tumor size reduction can be affected by other variations of the model such as the stress level-dependent switching rate, the growth of resistant tumor phenotype, and the intratumoral competition between sensitive and resistant phenotypes.   

\section{Therapy strategy I: Fixing the peak of sensitive tumor size during a single cycle of therapy}

In the stress-free period $t \in (0,t_f)$, we allow the sensitive tumor population of $S(t)$ to reach the fixed value of clinically tolerable tumor load, say $S_M$, during the stress-free time interval of each therapy cycle. For mathematical convenience, we choose $S_M=R_0 e^{\rho_s t_f}$ and $t_f=\frac{1}{\rho_s} ln(\frac{S_M}{R_0})$. We also use this value of $S_M$ as a normalization factor for both $S(t)$ and $R(t)$ terms and define the following terms: $\hat{r}_0=\frac{R_0}{S_M}$, $\hat{s}_1=\frac{S(t_f)}{S_M}$, and $\hat{r}_1=\frac{R(t_f)}{S_M}$. Using the defined $t_f$ and the normalized terms and Eqs~(\ref{S_Tf})-(\ref{R_Tf}), the normalized tumor sizes at $t=t_f$ can be expressed as follows: 
\begin{eqnarray}
\hat{s}_1 &=& 1+\hat{\beta} (1-\hat{r}_1) \\
\hat{r}_0 &=& (\hat{r}_1)^{1-\hat{\beta}} 
\label{r0_r1}
\end{eqnarray}
where we equate $\hat{\beta}=\frac{\beta}{\beta+\rho_s}$. Note that both $\hat{s}_1$ and $\hat{r}_0$ depends solely on the value of $\hat{r}_1$. 

In the stessful period, $t \in (t_f, t_f+t_s)$, we equate the sensitive and tolerant cell sizes at $t=t_f+t_s$, i.e., $S(t_f+t_s)=R(t_f+t_s)$. 
By normalizing all variables by $S_M$ and by letting $\hat{r}_2=\frac{R(t_f+t_s)}{S_M}$, we find the stressful time period, $t_s=\frac{1}{\rho_{eff}} ln \Big (\frac{(1+\hat{\alpha})(1+\hat{\beta}(1-\hat{r}_1))}{\hat{r}_1 + \hat{\alpha}(1+\hat{\beta}(1-\hat{r}_1))} \Big)$. From the Eqs.~(\ref{S_Ts})-(\ref{R_Ts}), the relationship between $\hat{r}_1$ and $\hat{r}_2$ can be expressed as follows:
\begin{equation}
\hat{r}_2= \frac{\hat{r}_1 (1-\hat{\alpha}\hat{\beta})+\hat{\alpha}(1+\hat{\beta})}{1+\hat{\alpha}}
\label{r2_r1}
\end{equation}
The time evolution of two tumor phenotypes during single cycle therapy is summarized in the Figure 1.

\section{Tumor size reduction criterion for one single cycle of therapy}
The tumor size reduction after one cycle of therapy requires that the total tumor size at the beginning of the cycle, $2R_0$, should be larger than the total tumor size at the ending of the cycle, $2R(t_f+t_s)$. This tumor reduction criterion can be simply stated as $\hat{r}_0 > \hat{r}_2$:
\begin{equation}
(\hat{r}_1)^{1-\hat{\beta}} > \frac{\hat{r}_1(1-\hat{\alpha}\hat{\beta})+\hat{\alpha}(1+\hat{\beta})}{1+\hat{\alpha}}
\end{equation}
The solution $\hat{r}_1$ that satisfies the above inequality can be found graphically. Firstly, $\hat{r}_2$ in Eq.~(\ref{r2_r1}) is a linear equation with the slope of $\frac{1-\hat{\alpha} \hat{\beta}}{1+\hat{\alpha}}$ and with the y-intercept of $\frac{\hat{\alpha}(1+\hat{\beta})}{1+\hat{\alpha}}$. Secondly, $\hat{r}_0$ in Eq.~(\ref{r0_r1}) has a hyperbolic shape because $0< 1-\hat{\beta}<1$ and $\hat{r}_0(\hat{r}_1=0)=0$ and $\hat{r}_0(\hat{r}_1=1)=1$. Now, two graphs of $\hat{r}_0$ and $\hat{r}_2$ intersect always at $\hat{r}_1=1$ and could potentially meet at another point at $\hat{r}^{*}_1$ where $0< \hat{r}^*_1<1$. To satisfy the tumor reduction criterion, it is necessary that two graphs should meet at the second point of $\hat{r}^*_1$ and $\hat{r}_1$ is chosen in the interval of $(\hat{r}^*_1,1)$. For the existence of the second intersection point of $\hat{r}^*_1$, the slope of the tangent line of $\hat{r}_0$ at $\hat{r}_1=1$, $\frac{d (\hat{r}_1)^{1-\hat{\beta}}}{d \hat{r}_1}|_{\hat{r}_1=1}=1-\hat{\beta}$, must be smaller than the slope of the linear equation of $\hat{r}_2$, $\frac{1-\hat{\alpha} \hat{\beta}}{1+\hat{\alpha}}$. This inequality condition of $1-\hat{\beta} < \frac{1-\hat{\alpha} \hat{\beta}}{1+\hat{\alpha}}$, can be further simplified as follows: 
\begin{equation}
\hat{\alpha} < \hat{\beta} \Longleftrightarrow \frac{\alpha}{\alpha+\delta-\rho_s} < \frac{\beta}{\beta+\rho_s}
\label{tumorreductioncriterion}
\end{equation}  
Therefore, the tumor reduction criteria are that (a) the model parameters satisfy the above inequality and $\hat{r}_1$ is selected from the interval of $(\hat{r}^*_1,1)$.

\section{Maximization of tumor size reduction after one cycle of therapy}
We intend to find the argument $\hat{r}_1$ that maximizes the difference between the initial tumor size $\hat{r}_0=(\hat{r}_1)^{1-\hat{\beta}}$ and the final tumor size $\hat{r}_2=\frac{\hat{r}_1(1-\hat{\alpha}\hat{\beta})+\hat{\alpha}(1+\hat{\beta})}{1+\hat{\alpha}}$. Let us define the tumor reduction function: $F(\hat{r}_1)=(\hat{r}_1)^{1-\hat{\beta}}- 
\frac{\hat{r}_1(1-\hat{\alpha}\hat{\beta})+\hat{\alpha}(1+\hat{\beta})}{1+\hat{\alpha}}$. The tumor reduction function $F(\hat{r}_1)$ is of a concave shape and has two roots, 1 and $\hat{r}^{*}_1$, if $\hat{\alpha} < \hat{\beta}$ and only one root, 1, otherwise. For $\hat{\alpha}=0$ and $\beta > 0$, $F(\hat{r}_1)$ always has two roots, $0$ and $1$. This is the case where only one phenotype of sensitive tumor exists and the tumor size always decreases after one cycle of therapy. For $\hat{\alpha} > 0$ and $\beta=0$, the aforementioned tumor reduction criterion is violated and $F(\hat{r}_1)$ has only one root, $1$. This case represents the irreversible switching of tumor cells to a resistant phenotype and tumor size always increases after one cycle of therapy. Now, if $\hat{\alpha}<\hat{\beta}$, the extremum of $F(\hat{r}_1)$ takes place at $\hat{r}^{**}_1$, and can be identified by setting $\frac{dF(\hat{r}_1)}{d\hat{r}_1}|_{\hat{r}^{**}_1}=0$. The treatment strategy becomes optimal, i.e., the tumor reduction after one cycle of therapy is maximized, when $\hat{r}_1=\hat{r}^{**}_1$:
\begin{equation}
\hat{r}^{**}_1=\Big ( 
\frac{(1-\hat{\beta})(\hat{\alpha}+1)}{1-\hat{\alpha} \hat{\beta}}
\Big )^{1/\hat{\beta}}
\end{equation}
where $\hat{r}^*_1 < \hat{r}^{**}_1 < 1$ because $1-\hat{\beta} < \frac{1-\hat{\alpha} \hat{\beta}}{1+\hat{\alpha}}$ and $\hat{\beta}<1$.

\section{Theorem 1: Limiting behavior of the tumor size after multiple cycles of therapy for the therapy strategy I}

{\bf Theorem 1:} When the tumor reduction criteion is met, i.e., $\hat{\alpha} < \hat{\beta}$, the total tumor size aymptotically approaches to $2 \hat{r}_{2N} \rightarrow 2(\hat{r}^*)^{1-\hat{\beta}}$ where $\hat{r}^*$ is the fixed point of the iterated map and $N$ is the number of therapeutic cycles. 
 
{\bf Proof of Theorem 1:}
During each stress-free period, we allow the sensitive tumor size to grow and reach the clinically tolerable tumor size, called $S_M(1+\hat{\beta}(1+\hat{r}_{2j+1}))$ for the $(j+1)$th cycle, and $S_M$ is set to be constant throughout the multiple cycles of therapy. All tumor sizes are normalized by $S_M$: $
\hat{s}_i=\frac{S_i}{S_M}$ and $\hat{r}_i=\frac{R_i}{S_M}$ where $\hat{r}_i$ and $\hat{s}_i$ are the normalized resistant and sensitive tumor sizes. the subscript $i$ specifies three time points during each cycle of therapy: $i=2j$ for the beginning, $i=2j+1$ for the intermediate when $\hat{s}_{2j+1}=1+\hat{\beta}(1+\hat{r}_{2j+1})$, and $i=2(j+1)$ for the ending time point of each cycle and $j=0,1,2... N-1$. The $(j+1)$th cycle of therapy starts with $\hat{s}_{2j} = \hat{r}_{2j}$, stops when $\hat{s}_{2(j+1)} = \hat{r}_{2(j+1)}$ and the next cycle starts over again. Imagine a sequence of $\hat{r}_i$ for N cycles of therapy: ($\hat{r}_0$, $\hat{r}_1$, $\hat{r}_2$), ($\hat{r}_2$, $\hat{r}_3$, $\hat{r}_4$), ...,($\hat{r}_{2(N-1)}$, $\hat{r}_{2N-1}$, $\hat{r}_{2N}$) where a single tuple represents the normalized resistant tumor sizes at three time points during each cycle of therapy. The time evolution of the normalized tumor sizes through the multiple cycles of therapy is presented in Figure 2. This progression of tumor size can be represented by the iterated map whose  updating rules are as follows: for $j=0,1,2,..., N-1$,
\begin{eqnarray}
\hat{r}_{2j} &\longrightarrow& \hat{r}_{2j+1} = (\hat{r}_{2j})^{\frac{1}{1-\hat{\beta}}} \\
\hat{r}_{2j+1} &\longrightarrow& \hat{r}_{2j+2}=\frac{\hat{r}_{2j+1}(1-\hat{\alpha}\hat{\beta})+\hat{\alpha}(1+\hat{\beta})}{1+\hat{\alpha}}
\end{eqnarray}
where during the $(j+1)$th cycle, the rest time period is $t^{(j+1)}_f=-\frac{1}{\rho_s}ln(\hat{r}_{2j})$ and the stressful time period is $t^{(j+1)}_s=\frac{1}{\rho_{eff}} ln \Big (\frac{(1+\hat{\alpha})(1+\hat{\beta}(1-\hat{r}_{2j+1}))}{\hat{r}_{2j+1} + \hat{\alpha}(1+\hat{\beta}(1-\hat{r}_{2j+1}))} \Big)$. 

There are two possible scenarios for the progression of tumor sizes: (i) when $\hat{\alpha} < \hat{\beta}$, the tumor reduction criterion is met and the resistant phenotype size converges to a fixed point smaller than $S_M$, (ii) when $\hat{\alpha} > \hat{\beta}$, the tumor reduction criterion is violated and the resistant phenotype size grows to another fixed point equal to $S_M$. First, we consider the case when the tumor reduction criterion is met, i.e., $\hat{\alpha} < \hat{\beta}$. In this case, there exists $\hat{r}^*$ such that $F(\hat{r}^*)=0$ and $0 \leq \hat{r}^* < 1$.  Suppose that given the initial resistant tumor size of $\hat{r}_0$, $\hat{r}_1=(\hat{r}_0)^{\frac{1}{1-\hat{\beta}}}$ is chosen within the interval of $(\hat{r}^*,1)$. As shown in Figure 3(a), given the value of the initial tumor size $\hat{r}_0$, $\hat{r}_1$ is determined by the graph of $\hat{r}_1=(\hat{r}_0)^{\frac{1}{1-\hat{\beta}}}$. Given the value of $\hat{r}_1$, $\hat{r}_2$ is determined by the graph of $\hat{r}_2=\frac{\hat{r}_1(1-\hat{\alpha}\hat{\beta})+\hat{\alpha}(1+\hat{\beta})}{1+\hat{\alpha}}$. During the subsequent cycles of therapy, $\hat{r}_{2j}$ determines $\hat{r}_{(2j+1)}$ from the graph of $\hat{r}_{2j+1}=(\hat{r}_{2j})^{\frac{1}{1-\hat{\beta}}}$ and $\hat{r}_{2j+1}$ determines $\hat{r}_{2j+2}$ from the graph of $\hat{r}_{2j+2}=\frac{\hat{r}_{2j+1}(1-\hat{\alpha}\hat{\beta})+\hat{\alpha}(1+\hat{\beta})}{1+\hat{\alpha}}$. This iterated map continues on for each successive cycle of therapy and the inequality relationship among $\hat{r}_{2j}$ for $j=0,1,2,...,N-1$ is given as $(\hat{r}^*)^{1-\hat{\beta}} <\hat{r}_{2N} < \hat{r}_{2N-2} < ... < \hat{r}_2 < \hat{r}_0 < 1$. In the limit of large $N$, the solution of the normalized resistant tumor size of $\hat{r}_{2N}$ converges to $(\hat{r}^*)^{1-\hat{\beta}}$ and thus the final total normalized tumor size after the $N$th cycle converges from above as follows:
\begin{equation} 
\hat{r}_{2N} + \hat{s}_{2N} \longrightarrow 2 (\hat{r}^*)^{1-\hat{\beta}}
\end{equation}

Suppose that given the initial tumor size $\hat{r}_0$, $\hat{r}_1=(\hat{r}_0)^{\frac{1}{1-\hat{\beta}}}$ is chosen outside of the interval $(\hat{r}^*,1)$, i.e., $\hat{r}_1 < \hat{r}^*$. Then, the value of the initial tumor size $\hat{r}_1$ is determined by the graph of $\hat{r}_1=(\hat{r}_0)^{\frac{1}{1-\hat{\beta}}}$ and subsequently $\hat{r}_2$ is determined by the graph of $\hat{r}_2=\frac{\hat{r}_1(1-\hat{\alpha}\hat{\beta})+\hat{\alpha}(1+\hat{\beta})}{1+\hat{\alpha}}$. However, since one of the tumor reduction criteria is not satisfied, the resistant tumor size at the end of subsequent cycles would increase rather than decrease, but still remain smaller than $(\hat{r}^*)^{1-\hat{\beta}}$ after one cycle of therapy, i.e., $\hat{r}_0 < \hat{r}_2 < (\hat{r}^*)^{1-\hat{\beta}}$. By repeating the same argument for the subsequent cycles, we can easily prove that the relationship of the normalized tumor sizes are given as follows: $\hat{r}_0 < \hat{r}_2 < \hat{r}_4 < ... <\hat{r}_{2N-2} < \hat{r}_{2N} < (\hat{r}^*)^{1-\hat{\beta}}$. This is depicted by the Figure 3(b). In the limit of large $N$, after $N$th cycle, the final total normalized tumor size of $\hat{s}_{2N} + \hat{r}_{2N}$  converges to $2(\hat{r}^*)^{1-\hat{\beta}}$ from below. As far as the tumor reduction criterion of $\hat{\alpha}<\hat{\beta}$ is satisfied, the resistant tumor size of $\hat{r}_{2j+2}$ after the $(j+1)$th cycle, converges to the same asymptotic value of $(\hat{r}^*)^{1-\hat{\beta}}$ regardless of the initial conditions, either $\hat{r}_0>\hat{r}^*$ or $ \hat{r}_0 <\hat{r}^*$. This indicates that the steady state solution $\hat{r}^*$ of the tumor reduction function $F(\hat{r})=0$ is a stable fixed point of the iterated map as shown in Figure 4(a).

Next, we consider the case when the tumor reduction criterion is violated, i.e., $\hat{\alpha} > \hat{\beta}$. Two graphs of $\hat{r}_0 (\hat{r}_1)$ and $\hat{r}_2(\hat{r}_1)$ intersect only at a single point of $\hat{r}_1=1$. As shown in Figure 5,  the iterated map leads us to the following sequence of the normalized resistant tumor sizes: $\hat{r}_0 < \hat{r}_1 < \hat{r}_2 < \hat{r}_3 ...$. In the limit of the large number $N$ of therapeutic cycles, the normalized sensitive and resistant tumor size converges to 1, i.e, $\hat{s}_{2N}$ and $\hat{r}_{2N} \rightarrow 1$, and the total tumor size converges to $\hat{s}_{2N}+\hat{r}_{2N} \rightarrow 2$. When the tumor reduction criterion is violated, the tumor reduction function $F(\hat{r}_1)$ has one stable fixed point of $\hat{r}=1$ as shown in Figure 4(b).

\section{Theorem 2: Asymptotically diminishing behavior of the tumor size for therapy strategy II utilizing the maximal tumor size reduction}

{\bf Theorem 2:}  When the tumor reduction criterion is met, i.e., $\hat{\alpha} < \hat{\beta}$ and $\hat{r}_{2j+1}$ is chosen to be $\hat{r}^{**}$ so that tumor size reduction is maximal after each cycle of therapy, the total tumor size asymptotically diminishes in the limit of large number of cycles where $\hat{r}^{**}$ maximizes the tumor size reduction function $F(\hat{r})$.

{\bf Proof of Theorem 2:} Therapy strategy II is to utilize the existence of the optimal resistant tumor size of $\hat{r}^{**}$ that maximizes the tumor reduction function $F(\hat{r})$ for the $(j+1)$th cycle and to always bring the normalized resistant tumor size to that optimal resistant tumor size of $\hat{r}^{**}$ at the designated mid time points for each $(j+1)$th therapy cycle. Consider a sequence of the normalized resistant tumor sizes during $N$ cycles of therapy: ($\hat{r}_0$, $\hat{r}_1$, $\hat{r}_2$), ($\hat{r}_2$, $\hat{r}_3$, $\hat{r}_4$), ...,($\hat{r}_{2(N-1)}$, $\hat{r}_{2N-1}$, $\hat{r}_{2N}$). Three normalized variables within the $(j+1)$th tuple indicates the resistant tumor sizes normalized by $S^{2j+1}_M$ at three different time points, $2j$, $2j+1$, and $2j+2$, for the $(j+1)$th cycle of therapy: $\hat{r}_{2j}=R_{2j}/S^{2j+1}_M$, $\hat{r}_{2j+1}=R_{2j+1}/S^{2j+1}_M$, and $\hat{r}_{2j+2}=R_{2j+2}/S^{2j+1}_M$. Due to the deliberate choice of the fixed value of $\hat{r}_{2j+1}=\hat{r}^{**}$ that maximizes tumor reduction size after a single cycle, the other two normalized tumor sizes within the same tuple are also fixed to the constant values, i.e., $\hat{r}_{2j}=(\hat{r}^{**})^{1-\hat{\beta}}$ and $\hat{r}_{2j+2}=\frac{\hat{r}^{**}+\hat{\alpha}(1+\hat{\beta}(1-\hat{r}^{**}))}{1+\hat{\alpha}}$. In other words, the sequence of the normalized resistant tumor sizes during $N$ number of cycles of therapy is simply represented by the $N$ times repetition of the same tuple of $((\hat{r}^{**})^{1-\hat{\beta}}, \hat{r}^{**}, \frac{\hat{r}^{**}+\hat{\alpha}(1+\hat{\beta}(1-\hat{r}^{**}))}{1+\hat{\alpha}})$. However, in contrast to the constancy of three normalized resistant tumor sizes within each tuple, the absolute values of resistant tumor sizes change over cycles of therapy due to the time-varying nature of the normalization factor, $S^{2j+1}_M(t)$. For example, for the first cycle, 
\begin{eqnarray}
R(0) &=& R_0=S^{(1)}_M (\hat{r}^{**})^{1-\hat{\beta}} \\
R(t_f) &=& R_1=S^{(1)}_M \hat{r}^{**} \\
S(t_f) &=& S_1=S^{(1)}_M (1+\hat{\beta}(1-\hat{r}^{**})) \\
R(t_f+t_s) &=& R_2=S^{(1)}_M \big( \frac{\hat{r}^{**}+\hat{\alpha}(1+\hat{\beta}(1-\hat{r}^{**}))}{1+\hat{\alpha}} \big) 
\end{eqnarray}
and for the second cycle, 
\begin{eqnarray}
R(t_f+t_s) &=& R_2=S^{(3)}_M (\hat{r}^{**})^{1-\hat{\beta}} \\
R(2t_f+t_2) &=& R_3=S^{(3)}_M \hat{r}^{**} \\
S(2t_f+t_s) &=& S_3=S^{(3)}_M (1+\hat{\beta}(1-\hat{r}^{**})) \\
R(2t_f+2t_s) &=& R_4=S^{(3)}_M \big( \frac{\hat{r}^{**}+\hat{\alpha}(1+\hat{\beta}(1-\hat{r}^{**}))}{1+\hat{\alpha}} \big) 
\end{eqnarray}
where the relationship between $S^{(1)}_M$ and $S^{(3)}_M$ is established at the overlapping time point between two consecutive cycles, say at $t=t_f+t_s$: $S^{(3)}_M =R_2 (\frac{1}{\hat{r}^{**}})^{1-\hat{\beta}}=S^{(1)}_M \big( \frac{\hat{r}^{**}+\hat{\alpha}(1+\hat{\beta}(1-\hat{r}^{**}))}{1+\hat{\alpha}} \big) (\frac{1}{\hat{r}^{**}})^{1-\hat{\beta}}$. In general, for the $(j+1)$th cycle, the normalization factor $S^{(2j+1)}_M$ is given as 
\begin{eqnarray}
\label{S_M}
S^{(2j+1)}_M &=& R_{2j} \Big( \frac{1}{\hat{r}^{**}} \Big)^{1-\hat{\beta}} \\ \nonumber
&=& S^{(2j-1)}_M \big( \frac{\hat{r}^{**}+\hat{\alpha}(1+\hat{\beta}(1-\hat{r}^{**}))}{1+\hat{\alpha}} \big) \big( \frac{1}{\hat{r}^{**}} \big)^{1-\hat{\beta}}
\end{eqnarray}

As shown in Eq. (\ref{S_M}), $S^{(2j+1)}_M$ is proportional to the absolute resistant tumor size of $R_{2j}$ at the beginning of the $(j+1)$th cycle and decreases after each successive cycle because of the maximal reduction of resistant tumor size per cycle, i.e.,  $R_{2j} > R_{2j+2}$ for $j=0,1,2,...N-1$. For the therapy strategy I, we control the normalization factor $S_M$ to be a "constant" value throughout the therapy cycles and the normalized sensitive tumor size grows to $S_M (1+\hat{\beta}(1-\hat{r}_{2j+1}))$ during the stress-free environment for the $(j+1)$th cycle. However for the therapy strategy II with the maximal tumor size reduction per each cycle, we control the normalized resistant tumor size of $\hat{r}_{2j+1}$ to reach the optimal normalized resistant tumor size of $\hat{r}^{**}$ and the normalized sensitive tumor size grows to a constant value of $S_M (1+\hat{\beta}(1-\hat{r}^{**}))$ during the stress-free environment for all cycles. 

To determine the rest period $t_f$ for the $(j+1)$th cycle, we solve the following two equations: 
$S^{(2j+1)}_M=R_{2j}e^{\rho_s t_f}$ and $R_{2j+1}=R_{2j} e^{-\beta t_f}$ which are normalized by $S^{(2j+1)}_M$ and are simplified further to $1=\hat{r}_{2j} e^{\rho_s t_f}$ and $\hat{r}^{**}=\hat{r}_{2j}e^{-\beta t_f}$.  Using $\hat{r}^{**}=(\frac{(1-\hat{\beta})(1+\hat{\alpha})}{1-\hat{\alpha}\hat{\beta}})^{\frac{1}{\hat{\beta}}}$, we obtain the expression for the rest period $t_f$, which is independent of the resistant tumor size of $\hat{r}_{2j}$ and remain constant throughout the multiple cycles of therapy: 
\begin{equation}
t_f = \frac{1}{\beta}ln \Big( \frac{1-\hat{\alpha}\hat{\beta}}{(1-\hat{\beta})(1+\hat{\alpha})}
\Big) 
\end{equation} 
This is in contrast to the therapy strategy I where $t_f=\frac{1}{\rho_s}ln(\frac{1}{\hat{r}_{2j}})$ depends on the normalized resistant tumor size at the beginning of each cycle and keeps increasing as the number of cycles increases when the tumor size reduction criterion is satisfied. 

To determine the stressful period $t_s$, we utilize the known expression for the normalized resistant tumor sizes during any stressful period: $\hat{r}_{2j+1}=\hat{r}^{**}$, $\hat{s}_{2j+1}=1+\hat{\beta}(1-\hat{r}^{**})$, and $\hat{r}_{2j+2}=\frac{\hat{r}^{**}+\hat{\alpha}(1+\hat{\beta}(1-\hat{r}^{**}))}{1+\hat{\alpha}}$. The stressful period $t_s$ is provided as 
\begin{eqnarray}
t_s &=& \frac{1}{\rho_{eff}}ln \Big( \frac{\hat{s}_{2j+1}}{\hat{r}_{2j+2}} \Big) \\ \nonumber
&=&\frac{1}{\rho_{eff}} ln \Big (\frac{(1+\hat{\alpha})(1+\hat{\beta}(1-\hat{r}^{**}))}{\hat{r}^{**} + \hat{\alpha}(1+\hat{\beta}(1-\hat{r}^{**}))} \Big)
\end{eqnarray}
Note that the stressful period $t_s$ is constant throughout the therapy cycles. 

From the information above, we find the iterated map for the progression of the resistant tumor sizes for the therapy strategy II: $R_{2j} \rightarrow S^{(2j+1)}_M=R_{2j}(\frac{1}{\hat{r}^{**}})^{1-\hat{\beta}} 
\rightarrow R_{2j+2}=S^{(2j+1)}_M \big( \frac{\hat{r}^{**}+\hat{\alpha}(1+\hat{\beta}(1-\hat{r}^{**}))}{1+\hat{\alpha}} \big)$. Iteratively applying the above map for the $N$ multiple cycles, we obtain
\begin{widetext}
\begin{eqnarray}
S^{(2N-1)}_M &=& R_{0} \Big( \frac{1}{\hat{r}^{**}} \Big)^{1-\hat{\beta}} 
\Big[ \Big( \frac{1}{\hat{r}^{**}} \Big)^{1-\hat{\beta}} 
\Big( \frac{\hat{r}^{**}+\hat{\alpha}(1+\hat{\beta}(1-\hat{r}^{**}))}{1+\hat{\alpha}} \Big)
\Big]^{N-1} \\
R_N &=& R_{0}  
\Big[ \Big( \frac{1}{\hat{r}^{**}} \Big)^{1-\hat{\beta}}
\Big( \frac{\hat{r}^{**}+\hat{\alpha}(1+\hat{\beta}(1-\hat{r}^{**}))}{1+\hat{\alpha}} \Big)
\Big]^{N} 
\end{eqnarray}
\end{widetext}
By the design of the maximal tumor reduction strategy, the tumor reduction function is maximized at the value of $\hat{r}^{**}$ and is positive, i.e., $F(\hat{r}^{**})=(\hat{r}^{**})^{1-\hat{\beta}}- \frac{\hat{r}^{**}+\hat{\alpha}(1+\hat{\beta}(1-\hat{r}^{**}))}{1+\hat{\alpha}} >0$. This positivity of the tumor reduction function for the value of $\hat{r}^{**}$ ensures that the argument inside the big bracket of $R_{N}$ is less than 1, i..e, 
$\Big( \frac{1}{\hat{r}^{**}} \Big)^{1-\hat{\beta}} \Big( \frac{\hat{r}^{**}+\hat{\alpha}(1+\hat{\beta}(1-\hat{r}^{**}))}{1+\hat{\alpha}} \Big) <1$. This implies that $R_N = R_0 a^N$ and $S^{(2N-1)}_M=R_0  \big( \frac{1}{\hat{r}^{**}} \big)^{1-\hat{\beta}} a^{N-1}$ where $a<1$. As the number of therapy cycles $N \rightarrow \infty$, both $R_N$ and $S^{(2N-1)}_M$ asymptotically diminish, i.e., 
\begin{eqnarray}
S^{(2N-1)}_M &\longrightarrow& 0 \\
R_N &\longrightarrow& 0
\end{eqnarray} 

\section{Effects of therapeutic dose-dependent phenotype switching rate $\alpha(\delta)$ on tumor size reduction}
In the prior sections, we treated the phenotype switching rate $\alpha$ as a constant, independent of the magnitude of therapeutic dose $\delta$. In this section, we investigate the effect of the dose-dependent switching rate, $\alpha(\delta)$ on tumor size reduction. We rewrite the tumor reduction function $F(\hat{r}_{2j+1})$ for the $(j+1)$th cycle of therapy, 
\begin{eqnarray}
F(\hat{r}_{2j+1}) &=& \hat{r}_{2j} - \hat{r}_{2j+2} = (\hat{r}_{2j+1})^{1-\hat{\beta}} \\
&-& \frac{\hat{r}_{2j+1} (1-\hat{\alpha}(\delta)\hat{\beta}) +\hat{\alpha}(\delta)(1+\hat{\beta})}{1+\hat{\alpha}(\delta)}
\end{eqnarray} 
where $\hat{\alpha}(\delta) = \frac{\alpha(\delta)}{\alpha(\delta)+\delta -\rho_s}$. Below we consider two different cases of dose-dependency of the forward switching rate, linear and saturating cases.  

(i) Linear case: $\alpha(\delta)=c \delta \Theta(\delta-\rho_s)$. We assume that the forward switching rate linearly depends on the therapeutic dose and we require $\delta>\rho_s$ for the sufficient efficacy of therapy, i.e., $\alpha(\delta)=c \delta \Theta(\delta-\rho_s)$ where $c$ is a constant and $\Theta(x)$ is a Heaviside step function: $\Theta(x)=1$ for $x>0$ and otherwise $\Theta(x)=0$. 

{\bf Proposition 3:} For the linear case, if $\hat{\beta}>\frac{c}{c+1}$, tumor size reduction criterion is met for $\delta > \delta_c$, and tumor size converges to a constant value or diminish asymptotically after a large number of therapy cycles per theorem 1 and 2. However, if $\hat{\beta} < \frac{c}{c+1}$, tumor size reduction criterion cannot be met regardless of the value of $\delta$ and tumor size always increases. 

{\bf Proof of Proposition 3:} 
$\hat{\alpha}(\delta)$ is a decreasing function of the therapeutic dose; it starts from $\hat{\alpha}(\delta=\rho)=1$ and decreases to $\frac{c}{c+1}$ in the limit of a large value of $\delta$. The tumor reduction criterion in Eq.~(\ref{tumorreductioncriterion}) is rewritten again as a function of $\delta$,
\begin{equation}
\hat{\alpha}(\delta)< \hat{\beta} \Longleftrightarrow \frac{c\delta}{(c+1)\delta-\rho_s} < \hat{\beta}
\label{linearcase}
\end{equation}
If $\hat{\beta} > \frac{c}{c+1}$, the tumor reduction criterion is satisfied for a certain range of therapeutic dose $\delta> \delta_c$ where $\hat{\alpha}(\delta_c)=\hat{\beta}$ and $\delta_c=\frac{\hat{\beta}\rho}{\hat{\beta}(c+1)-c}$.  The above tumor size reduction criterion in Eq.~(\ref{linearcase}) is nothing but the comparison of two slopes at $\hat{r}_{2j+1}=1$, one from the nonlinear function of $\hat{r}_{2j}$ in Eq.~(\ref{r0_r1}) and another from the linear function of $\hat{r}_{2j+2}$ in Eq~(\ref{r2_r1}). The slope of a linear function of $\hat{r}_{2j+2}$, $\frac{1-\hat{\alpha}(\delta) \hat{\beta}}{1+\hat{\alpha}(\delta)}$, is a monotone decreasing function of $\hat{\alpha}$ and is a monotone increasing function of $\delta$ (because $\hat{\alpha}(\delta)$ is a monotone decreasing function of $\delta$). As $\delta$ increases, the slope of $\hat{r}_{2j}$ remain unchanged, but the slope of $\hat{r}_{2j+2}$ increases, which means that two functions intersect at the smaller value of $\hat{r}^*$ and thus the tumor reduction function $F(\hat{r}_{2j+1})$ increases. In words, as the therapeutic dose increases, the amount of tumor reduction after each cycle of therapy increases and after multiple cycles of therapy the tumor size asymptotically decreases to its limiting value. If $\hat{\beta} < \frac{c}{c+1}$, the tumor reduction criterion is violated for any value of therapeutic dose $\delta>\rho$ and tumor size will grow despite therapy. 

(ii) Saturating case: $\alpha(\delta)=\frac{c \delta \Theta(\delta-\rho_s)}{\delta + d}$. Here we assume that the forward switching rate takes a saturating functional form; $\alpha(\delta)$ increases and then saturates to a constant $c$ as $\delta$ increases. 

{\bf Proposition 4:} For the saturating case, tumor size decreases if $\delta>\delta_c$ regardless of $\hat{\beta}$, and tumor size converges to a constant value or diminishes asymptotically after a large number of therapy cycles per theorem 1 and 2. 

{\bf Proof of Proposition 4:} $\hat{\alpha}$ is again a decreasing function of the therapeutic dose; it starts from $\hat{\alpha}(\delta=\rho_s)=1$ and decreases to zero in the limit of a large value of $\delta$. Thus, for any value of $\hat{\beta}$, there exists $\delta>\delta_c$ such that the tumor reduction criterion of $\hat{\alpha}(\delta) < \hat{\beta}$ is satisfied and $\delta_c$ is defined as $\hat{\alpha}(\delta_c)=\hat{\beta}$. Like the linear case, the slope of a linear function of $\hat{r}_{2j}$, $\frac{1-\hat{\alpha}(\delta) \hat{\beta}}{1+\hat{\alpha}(\delta)}$, is a monotone increasing function of $\delta$, bounded between $\frac{1-\hat{\beta}}{2}$ and $1$. As $\delta$ increases, the slope of $\hat{r}_{2j}$ approaches one and its y-intercept, $\frac{\hat{\alpha}(\delta)(1+\hat{\beta})}{1+\hat{\alpha}(\delta)}$, goes to zero. Consequently, two curves of $\hat{r}_{2j-2}$ and $\hat{r}_{2j}$ intersect at two points, $\hat{r}^*_0=0$ and $1$ and the tumor reduction function becomes $F(\hat{r}_{2j-1})=(\hat{r}_{2j-1})^{1-\hat{\beta}}-\hat{r}_{2j-1}$ after each cycle. The maximal tumor reduction after each cycle takes place at $\hat{r}^{**}=(1-\hat{\beta})^{1/\hat{\beta}}$. In words, in this saturating case of the forward switching rate, the asymptotic limit of tumor size approaches to zero as the therapeutic dose increases. Thus, the optimal therapeutic strategy is to maximize the therapeutic dose during the stressful period.

\section{Effects of non-zero growth rate of tolerant phenotype on tumor size reduction}
We relax our strong assumption of the quiescence of the resistant phenotype and allow the resistant phenotypic cells to grow at a growth rate of $\rho_r$. We assume that the growth rate of sensitive phenotype is larger than that of resistant phenotype, i.e. $\rho_s>\rho_r$. The modified coupled ODE system is given for the stress-free/rest period, $t \in (0,t_f)$: 
\begin{eqnarray}
\frac{dS(t)}{dt} &=& \rho_s S + \beta R \\
\frac{d R(t)}{dt} &=& \rho_r R -\beta R
\end{eqnarray}
The solution of this system of coupled ODE is provided with the initial conditions of $S(0)=R(0)=R_0$.
\begin{eqnarray}
S(t) &=& R_0 e^{\rho_s t} +\tilde{\beta} R_0 (e^{\rho_s t} - e^{-(\beta -\rho_r) t}) \\
R(t) &=& R_0 e^{- (\beta -\rho_r) t}
\end{eqnarray}
where $\tilde{\beta}=\frac{\beta}{\rho_s-\rho_r+\beta}$. After normalization of the sensitive and resistant phenotype variables with a constant normalization factor of $S_M=S_0 e^{\rho_s t_f}$ as in the previous section, we recover the similar relationship among $\hat{r}_0$, $\hat{s}_1$, and $\hat{r}_1$ with the rest/stress-free period $t_f=-\frac{1}{\rho_s} ln (\hat{r}_0)$.  
\begin{eqnarray}
\hat{s}_1 &=& 1+\tilde{\beta}(1-\hat{r}_1) \\
\hat{r}_1 &=& (\hat{r}_0)^{\frac{\rho_s -\rho_r+\beta}{\rho_s}}
\label{r0_growth}
\end{eqnarray} 

For the stressful period $t \in (t_f, t_f+t_s)$, the modified ODE system is given
\begin{eqnarray}
\frac{d S(t)}{dt} &=& (\rho_s - \delta -\alpha) S \\
\frac{d R(t)}{dt} &=& \rho_r R + \alpha S
\end{eqnarray}
The solution of the above system is provided with the initial conditions of $S(t_f)$ and $R(t_f)$. 
\begin{eqnarray}
S(t) &=& S(t_f) e^{-\rho_{eff}(t-t_f)} \\
R(t) &=& R(t_f) e^{\rho_r (t-t_f)} +\tilde{\alpha} S(t_f) (e^{\rho_r (t-t_f)} - e^{-\rho_{eff} (t-t_f)})
\end{eqnarray}
where $\rho_{eff}=\delta+\alpha-\rho_s$ and $\tilde{\alpha}=\frac{\alpha}{\rho_r+\rho_{eff}}$. By normalizing both phenotype variables with $S_M$ and setting $S(t_f+t_s)=R(t_f+t_s)$ at $t=t_f+t_s$, we identify the required stressful period, $t_s$, that makes the population sizes of two phenotypes coincide again at $t=t_f+t_s$: $t_s=-\frac{1}{\rho_r+\rho_{eff}}ln \Big( \frac{\hat{r}_1 (1-\tilde{\alpha} \tilde{\beta}) +\tilde{\alpha} (1+\tilde{\beta})}{(1+\tilde{\alpha})(1+\tilde{\beta}(1-\hat{r}_1))} \Big)$ as well as the relationship between $\hat{r}_1$ and $\hat{r}_2$, 
\begin{eqnarray} \label{r2_growth}
\hat{r}_2 &=& \hat{s}_1 \Big( \frac{\hat{r}_1 +\tilde{\alpha} \hat{s}_1}{\hat{s}_1 (1+\tilde{\alpha})} \Big)^{\frac{\rho_{eff}}{\rho_r+\rho_{eff}}}
\\
&=& \Big( \frac{[1+\tilde{\beta}(1-\hat{r}_1)]^{\frac{\rho_r}{\rho_{eff}}}
(\hat{r}_1+\tilde{\alpha} [1+\tilde{\beta}(1-\hat{r}_1)])}
{1+\tilde{\alpha}} \Big)^{\frac{\rho_{eff}}{\rho_r+\rho_{eff}}} 
\label{r2_growth}
\end{eqnarray} 

As discussed in the previous section, the tumor reduction criterion after one cycle of therapy requires the following inequality condition: $F(\hat{r}_1) = \hat{r}_0(\hat{r}_1) -\hat{r}_2 (\hat{r}_1) >0$ where $0<\hat{r}_1<1$. The function $\hat{r}_2(\hat{r}_1)$ in Eq. (\ref{r2_growth}) is a monotone increasing function of $\hat{r}_1$, of a hyperbolic shape, and satisfies $\hat{r}_2(\hat{r}_1=0)=(1+\tilde{\beta})(\frac{\tilde{\alpha}}{\tilde{\alpha}+1})^{\frac{\rho_{eff}}{\rho_{eff}+\rho_r}}$ and $\hat{r}_2(\hat{r}_1=1)=1$. The function $\hat{r}_0$ in Eq. (\ref{r0_growth}) is a monotone increasing function of $\hat{r}_1$ and meets $\hat{r}_0(\hat{r}_1=0)=0$ and 
$\hat{r}_0(\hat{r}_1=1)=1$. Depending on the inequality relationship between $\beta$ and $\rho_r$, the shape of the function $\hat{r}_0(\hat{r}_1)$ can be of a hyperbolic shape or of a convex shape. (i) When $\beta<\rho_r$, $\frac{\rho_s}{\rho_s -\rho_r+\beta}>1$ and the function $\hat{r}_0$ is of a convex shape and intersects with the function $\hat{r}_2$ only at $\hat{r}_1=1$ in the range of $0<\hat{r}_1<1$. As shown in Fig. \ref{Fig8:iterativemap_growth}, after each $(j+1)$th cycle of therapy, the value of $\hat{r}_{2j+2}$ increases and asymptotically approaches to $\hat{r}_{2j+1} \rightarrow 1$ for a large $N$ number of cycles. (ii) When $\beta>\rho_r$, $\frac{\rho_s}{\rho_s -\rho_r+\beta}<1$ and the function $\hat{r}_0$ is of hyperbolic shape. It could potentially intersect with the function of $\hat{r}_2$ at the second intersection point, say $\hat{r}^*$, in addition to $\hat{r}_1=1$ where $0<\hat{r}^*_1<1$ if the following tumor reduction criterion is satisfied: $\frac{d \hat{r}_0}{d \hat{r}_1}|_{\hat{r}_1=1} < \frac{d \hat{r}_2}{d \hat{r}_1}|_{\hat{r}_1=1}$, or in other words,
\begin{equation}
\frac{\rho_s}{\rho_s-\rho_r +\beta} < \frac{\rho_{eff}}{\rho_{eff}+\rho_r} \Big( \frac{1-\tilde{\alpha}\tilde{\beta}}{1+\tilde{\alpha}}-\frac{\rho_r \tilde{\beta}}{\rho_{eff}} \Big)  
\end{equation}
After rearranging terms, the above tumor reduction criterion with the non-zero growth rate of resistant phenotype can be simplified as 
\begin{equation}
\tilde{\alpha} < \hat{\beta}+\tilde{\alpha}(\hat{\beta}-\tilde{\beta}) -G(\rho_r)
\label{tumorreductioncriterion_growth}
\end{equation}
where $\tilde{\alpha}=\frac{\alpha}{\rho_{eff}+\rho_r}$, $\hat{\alpha}=\frac{\alpha}{\rho_{eff}}$, $\tilde{\beta}=\frac{\beta}{\rho_s-\rho_r+\beta}$, $\hat{\beta}=\frac{\beta}{\rho_s+\beta}$, and $G(\rho_r)=\rho_r [(1+\tilde{\alpha})\frac{\tilde{\alpha}}{\alpha}+(1-\tilde{\alpha} \tilde{\beta})\frac{\tilde{\beta}}{\beta}]$. Since we assume $\rho_s>\rho_r$ and $\delta>\rho_s$ (the therapeutic killing rate being larger than the growth rate of sensitive phenotype for the therapeutic efficacy) and the positivity of all the model parameter values, $\tilde{\alpha}< 1$ and $\tilde{\beta}<1$ and thus $G(\rho_r)>0$ always. For the zero growth rate of resistant phenotype ($\rho_r=0$), $\tilde{\alpha}=\hat{\alpha}$ and $\tilde{\beta}=\hat{\beta}$ and we recovers the original tumor reduction criterion of $\hat{\alpha} < \hat{\beta}$ in Eq. (\ref{tumorreductioncriterion}). 

This non-zero growth rate makes the tumor reduction criterion more difficult to be satisfied as $\rho_r$ increases. This can be shown in the limit of small growth rate of $\rho_r$. Taylor-expanding the above tumor reduction criterion around $\rho_r=0$, we obtain the following relation:
\begin{equation}
\hat{\alpha} < \hat{\beta} -H(\rho_r) 
\label{tumorreduction_smallrho}
\end{equation}
where $H(\rho_r)=\rho_r(\frac{\hat{\alpha}}{\alpha}+\frac{\hat{\beta}}{\beta})-\rho^2_r [\hat{\alpha}(\frac{\hat{\alpha}}{\alpha})^2+\hat{\alpha}\hat{\beta} (\frac{\hat{\beta}}{\beta})^2] + \rho^3_r (\frac{\hat{\alpha}}{\alpha})(\frac{\hat{\beta}}{\beta})^3$. In the limit of small $\rho_r$, the first term in $H(\rho_r)$ dominates and decreases the right hand side of the Eq. (\ref{tumorreduction_smallrho}), making the tumor reduction criterion hard to be achieved. As far as the above tumor reduction criterion in Eq. (\ref{tumorreductioncriterion_growth}) is satisfied, the tumor size can be reduced to either $(\hat{r}^*)^{\frac{\rho_s}{\rho_s-\rho_r+\beta}}$ or zero after a large number of therapeutic cycles by following either of two therapy strategies outlined in the previous sections, the fixed $S_M$ scheme in the section (VII) or the maximal tumor reduction scheme in the section (VIII), respectively.  

\section{Effects of nonlinear tumor growth rate and intratumoral competition on tumor reduction} 

Instead of introducing the nonlinear growth rate to both sensitive and resistant tumor phenotypes, we let the growth of the resistant phenotype size to be regulated and suppressed by the growing sensitive phenotype size, but not the other way around. This choice of the asymmetric intratumoral competition is for mathematical convenience of the minimalist model that simulates the containment of the slowly growing resistant tumor phenotype through competition with the rapidly growing sensitive tumor phenotype over limited resources in a tumor microenvironment. In our model, this intratumoral competition with the sole intent of resistant tumor growth suppression is likely to compete with the non-zero growth rate of the resistant phenotype as well as the stress-induced phenotype switching of sensitive to resistant phenotype.   
During the rest/stress-free period, $t \in (0,t_f)$, 
\begin{eqnarray}
\frac{dS}{dt} &=& \rho_s S + \beta R \\
\frac{dR}{dt} &=&  \rho_r R(1-\epsilon S) -\beta R 
\end{eqnarray}

During the stressful  period, $t \in (t_f,t_f+t_s)$, 
\begin{eqnarray}
\frac{dS}{dt} &=&  -(\alpha + \delta - \rho_s) S \\
\frac{dR}{dt} &=&  \rho_r R(1-\epsilon S) + \alpha S -\delta_r R
\end{eqnarray}

We use the small $\epsilon$ perturbation technique. The following solution Ansatz for $S(t)$ and $R(t)$ are assumed in the limit of small $\epsilon$: $S(t)=\sum_{n=0} \epsilon^n S_n (t)$ and $R(t)=\sum_{n=0} \epsilon^n R_n (t)$. For the mathematical convenience, we simplify the Ansatz to $S(t)=S_1(t) + \epsilon S_2(t)$ and $R(t)=R_1(t) +\epsilon R_2(t)$. After substituting $S(t)$ and $R(t)$ with the simplified Ansatz, we obtain two sets of ODE, the first set in the order of $O(1)$ and the second set in order of $O(\epsilon)$. We will present the small $\epsilon$ perturbation solutions in the next sections.   

\subsection{Rest period: O(1) solution}
After collecting the terms in the order of $O(1)$, during the rest/stress-free period, $t \in (0,t_f)$, the set of ODE  for the time evolution of two variables, $S_1(t)$ and $R_1(t)$, is provided as below: 
\begin{eqnarray}
\frac{dS_1}{dt} &=& \rho_s S_1 + \beta R_1 \\
\frac{dR_1}{dt} &=&  (\rho_r - \beta) R_1 
\end{eqnarray}
Their solutions with the initial conditions of $S_1(0)=R_1(0)=R_0$ are $R_1(t)=R_0 e^{(\rho_r-\beta)t}$ and $S_1(t)=R_0 e^{\rho_s t} +\tilde{\beta} R_0 e^{\rho_s t} (1-e^{-(\beta +\rho_s- \rho_r)t})$ where $\tilde{\beta}=\frac{\beta}{\beta+\rho_s-\rho_r}$. As done for the therapy strategy I, We fix the sensitive tumor size to a constant value of $S_{M}$ at the end of the rest period, i.e., at $t=t_f$: $S_M=R_0 e^{\rho_s t_f}$. All of the tumor sizes are normalized by this constant normalization factor $S_M$: $r_0=\frac{R_0}{S_M}$, $\tilde{r}_1=\frac{R_1(t_f)}{S_M}$, and $\tilde{s}_1=\frac{S_1(t_f)}{S_M}$. Then, the rest period time is determined to be $t_f=\frac{1}{\rho_s} ln (\frac{1}{r_0})$. Those normalized tumor sizes at the beginning and the ending of the rest period are related as follows: 
\begin{eqnarray}
\tilde{r}_1 &=& r_0 e^{(\rho_r -\beta) t_f} = (r_0)^{\frac{\rho_s+\beta-\rho_r}{\rho_s}} \\
\tilde{s}_1 &=& 1+\tilde{\beta}(1-\tilde{r}_1)
\end{eqnarray}

\subsection{Rest period: $O(\epsilon)$ solution}
Collecting the terms in the order of $O(\epsilon)$, during the rest/stress-free period, $t \in (0,t_f)$, the set of ODE  for the time evolution of $S_2(t)$ and $R_2(t)$, is provided as below: 
\begin{eqnarray}
\frac{dS_2}{dt} &=& \rho_s S_2 + \beta R_2 \\
\frac{dR_2}{dt} &=&  (\rho_r - \beta) R_2 -\rho_r R_1(t) S_1(t) 
\end{eqnarray}
After moving the first two terms on the right hand side of Eq. (10) to its left hand side, the ODE for $R_2(t)$ is rewritten as follows: $\frac{d [R_2(t) u(t)]}{dt}=-\rho_r u(t) R_1(t) S_1(t)$ where $u(t)=e^{(\beta-\rho_r)t}$. We substitute $R_1(t)$ and $S_1(t)$ in the Eq. (10) with the $O(1)$ solution of the set of ODE in the Eqs. (6)-(7). The $O(\epsilon)$ solution for $R_2(t)$ is given as  
\begin{equation}
R_2(t) = \rho_r R_0 e^{(\rho_r-\beta)t}
\Big[
\frac{(1+\tilde{\beta})(R_0-R_0e^{\rho_s t})}{\rho_s}
+ \frac{\tilde{\beta} (R_0-R_0 e^{(\rho_r -\beta)t})}{\beta-\rho_r}
\Big]
\end{equation}

After moving the first term on the r.h.s to the l.h.s in the Eq. (9), we rewrite the ODE for $S_2(t)$ as follows: $\frac{d [S_2(t) w(t)]}{dt}= \beta w(t) R_2(t) $ where $w(t)=e^{-\rho_s t}$. Substituting $R_2(t)$ with its $O(\epsilon)$ solution in the Eq. (11), we solve for $S_2(t)$: 
\begin{eqnarray}
S_2(t) &=&
\beta \rho_r
\Big[
(\frac{1+\tilde{\beta}}{\rho_s}+\frac{\tilde{\beta}}{\beta-\rho_r})
\frac{R_0(R_0 e^{\rho_s t}-R_0 e^{(\rho_r-\beta)t})}{\beta+\rho_s-\rho_r} 
\\ \nonumber
&+& \frac{(1+\tilde{\beta})R_0 e^{\rho_s t} (R_0 e^{(\rho_r-\beta)t}-R_0)}{\rho_s(\beta-\rho_r)} 
+ \frac{\tilde{\beta}(R^2_0 e^{2(\rho_r - \beta) t}-R^2_0e^{\rho_s t})
}{(\beta-\rho_r)(\rho_s+2\beta-2\rho_r)} 
\Big]
\end{eqnarray}

At the end of the rest period, $t=t_f$, using $S_M=R_0 e^{\rho_s t_f}$ and $R_1(t_f)=R_0 e^{(\rho_r-\beta)t_f}$, the perturbative solutions of $S(t)$ and $R(t)$ up to the first order of $\epsilon$ are given as follows:  
\begin{eqnarray}
R(t_f) &=& R_1(t_f) +\epsilon R_2(t) 
\\ \nonumber
&=& R_1(t_f) \Big[
1+\epsilon \rho_r \big(
\frac{(1+\tilde{\beta})(R_0 - S_M)}{\rho_s}  + \frac{\tilde{\beta}(R_0-R_1(t_f))}{\beta-\rho_r} 
\big)
\Big] 
\\ 
S(t_f) &=& S_1(t_f) +\epsilon S_2(t) 
\\ \nonumber
&=& S_M+\tilde{\beta} (S_M -R_1(t_f))
+\epsilon \rho_r \beta 
\Big[ 
(\frac{1+\tilde{\beta}}{\rho_s}+\frac{\tilde{\beta}}{\beta-\rho_r}) 
\frac{R_0(S_M-R_1(t_f))}{\beta+\rho_s-\rho_r}
\\ \nonumber
&+& \frac{(1+\tilde{\beta}) S_M (R_1(t_f)-R_0) }{\rho_s(\beta-\rho_r)} 
+ \frac{\tilde{\beta} (R^2_1(t_f)-R_0 S_M)}{(\beta-\rho_r)(\rho_s+2\beta-2\rho_r)} 
\Big]
\end{eqnarray}

By normalizing all of the tumor sizes by the normalization factor $S_M$, we define the dimensionless parameters: $\tilde{\epsilon}=\epsilon  S_M$,   $r_0 =\frac{R_0}{S_M}$, $r_1=\frac{R(t_f)}{S_M}$, $s_1=\frac{S(t_f)}{S_M}$, and $\tilde{r}_1=\frac{R_1(t_f)}{S_M}$. 
\begin{eqnarray}
r_1 &=& \tilde{r}_1 \Big[
1+\tilde{\epsilon} \rho_r \big(
\frac{(1+\tilde{\beta})}{\rho_s} (r_0 -1) +\frac{\tilde{\beta}}{\beta-\rho_r}(r_0 -\tilde{r}_1)
\big)
\Big] 
\\ 
s_1 &=& 1+\tilde{\beta}(1-\tilde{r}_1) 
+\tilde{\epsilon} \rho_r \beta 
\Big[ 
(\frac{1+\tilde{\beta}}{\rho_s}+\frac{\tilde{\beta}}{\beta-\rho_r}) 
\frac{r_0 (1-\tilde{r}_1)}{\beta+\rho_s-\rho_r}
\\ \nonumber
&+& \frac{(1+\tilde{\beta}) (\tilde{r}_1 -r_0) }{\rho_s(\beta-\rho_r)} 
+ \frac{\tilde{\beta} (\tilde{r}^2_1 -r_0)}{(\beta-\rho_r)(\rho_s+2\beta-2\rho_r)} 
\Big]
\end{eqnarray} 

\subsection{Stressful Period: O(1) solution}
In the stressful period, $t \in (t_f,t_f+t_s)$, the ODE for $S(t)$ is decoupled from the ODE for $R(t)$. This enables us to find the full solution for $S(t)$ without the need of a perturbative solution. We only apply the simplified Ansatz to R(t) solution: $R(t)=R_1+\epsilon R_2$. After collecting the terms in the order of $O(1)$, the set of ODE for the time evolution of two variables, $S(t)$ and $R_1(t)$, is provided as below: 
\begin{eqnarray}
\frac{dS}{dt} &=& -\rho_{eff} S\\
\frac{dR_1}{dt} &=&  (\rho_r - \delta_r) R_1 +\alpha S 
\end{eqnarray}
where $\rho_{eff}=\delta_s+\alpha-\rho_s$. The ODE for $S(t)$ is decoupled from the ODE for $R_1(t)$ and its solution with the initial condition of $S(t_f)$ is $S(t)=S(t_f) e^{-\rho_{eff}(t-t_f)}$. By moving the first two terms from the r.h.s to the l.h.s in Eq. (16), we rewrite the ODE for $R_1(t)$ as follows: $\frac{d [R_1(t) x(t)]}{dt}= \alpha x(t) S_1(t)$ where $x(t)=e^{(\delta_r-\rho_r) t}$. Substituting $S(t)$ with its full solution, we solve for $R_1(t)$. Below is the O(1) solution of $R_1(t)$:  
\begin{equation}
R_1(t) = ( R(t_f) +\tilde{\alpha} S(t_f)) e^{(\rho_r-\delta_r)(t-t_f)} 
- \tilde{\alpha} S(t_f) e^{-\rho_{eff}(t-t_f)}
\end{equation}
where $\tilde{\alpha}=\frac{\alpha}{\rho_{eff}+\rho_r-\delta_r}$ and $S(t_f)$ and $R(t_f)$ are the tumor sizes at the end of the stress-free period as defined as in the Eqs. (13) and (14). All of the tumor sizes are normalized by the constant normalization factor $S_M$: $s_1=\frac{S(t_f)}{S_M}$, $r_1=\frac{R(t_f)}{S_M}$, and $\tilde{r}_2=\frac{R_1(t_f+t_s)}{S_M}$. Those normalized tumor sizes at the beginning and the ending of the stressful period are related as follows: 
\begin{equation}
\tilde{r}_2  = (r_1 +\tilde{\alpha} s_1)e^{(\rho_r-\delta_r)t_s}- \tilde{\alpha} s_1 e^{-\rho_{eff} t_s}
\end{equation}

\subsection{Stressful period: $O(\epsilon)$ solution}
We solve for $O(\epsilon)$ solution of $R_2(t)$ for the stressful period, $t \in (t_f,t_f+t_s)$. Collecting the terms in the order of $O(\epsilon)$, the ODE for the time evolution of $R_2(t)$ is provided as below: 
\begin{equation}
\frac{dR_2}{dt} =  (\rho_r - \delta_r) R_2 -\rho_r S R_1  
\end{equation}
After moving the first two terms on the right hand side of Eq. (23) to its left hand side, the ODE for $R_2(t)$ is rewritten as follows: $\frac{d [R_2(t)z(t)]}{dt}=-\rho_r z(t) S(t)R_1(t)$ where $z(t)=e^{(\beta-\rho_r)t}$. We substitute $S(t)$ and $R_1(t)$ with their full and $O(1)$ solutions respectively. The $O(\epsilon)$ solution for $R_2(t)$ is given as  
\begin{eqnarray}
R_2(t) &=& \rho_r S(t_f)e^{-\rho_{eff}(t-t_f)} 
\Big[
\big( \frac{R(t_f)+\tilde{\alpha} S(t_f)}{\rho_{eff}} \big)
(e^{(\rho_r-\delta_r)(t-t_f)}-e^{(\rho_{eff}+\rho_r-\delta_r)(t-t_f)}) \\ \nonumber
&+& \frac{\tilde{\alpha} S(t_f)e^{-\rho_{eff}(t-t_f)} (e^{(2\rho_{eff}+\rho_r-\delta_r)(t-t_f)}-1)}{2 \rho_{eff}+\rho_r-\delta_r}
\Big]
\end{eqnarray}

At the end of the stressful period, $t=t_f+t_s$, the full solution of $S(t)$ and the perturbative solution of $R(t)$ up to the first order of $\epsilon$ are given as follows: 
\begin{eqnarray}
S(t_f+t_s) &=& S(t_f) e^{-\rho_{eff} t_s} \\
R(t_f+t_s) &=& R_1(t_f+t_s)+\epsilon R_2(t_f+t_s) \\ \nonumber
&=& (R(t_f) +\tilde{\alpha} S(t_f)) e^{(\rho_r-\delta_r)t_s}  
- \tilde{\alpha} S(t_f)  e^{-\rho_{eff}t_s} 
\\ \nonumber
&+& \epsilon \rho_r S(t_f)e^{-\rho_{eff}t_s} 
\Big[
\big( \frac{R(t_f)+\tilde{\alpha} S(t_f)}{\rho_{eff}} \big)
(e^{(\rho_r-\delta_r)t_s}-e^{(\rho_{eff}+\rho_r-\delta_r)t_s}) 
\\ \nonumber
&+& \frac{\tilde{\alpha} S(t_f) (e^{(\rho_{eff}+\rho_r-\delta_r)t_s}-e^{-\rho_{eff}t_s})}{2 \rho_{eff}+\rho_r-\delta_r}
\Big]
\end{eqnarray}

By normalizing all of the tumor sizes by the normalization factor $S_M$, we define the dimensionless parameters: $\tilde{\epsilon}=\epsilon  S_M$,   $r_0 =\frac{R_0}{S_M}$, $r_1=\frac{R(t_f)}{S_M}$, $s_1=\frac{S(t_f)}{S_M}$, $\tilde{r}_1=\frac{R_1(t_f)}{S_M}$, $s_2=\frac{S(t_f+t_s)}{S_M}$, and $r_2=\frac{R(t_f+t_s)}{S_M}$:
\begin{eqnarray}
s_2 &=& s_1 e^{-\rho_{eff} t_s}  \\
r_2 &=& (r_1 +\tilde{\alpha} s_1) e^{(\rho_r -\delta_r) t_s} 
-\tilde{\alpha} s_1 e^{-\rho_{eff} t_s} \\
&+& \tilde{\epsilon} \rho_r s_1 e^{-\rho_{eff} t_s} 
\Big[
\big( \frac{r_1+\tilde{\alpha} s_1}{\rho_{eff}} \big)
(e^{(\rho_r-\delta_r)t_s}-e^{(\rho_{eff}+\rho_r-\delta_r)t_s}) 
\\ \nonumber
&+& \frac{\tilde{\alpha} s_1  (e^{(\rho_{eff}+\rho_r-\delta_r)t_s}-e^{-\rho_{eff}t})}{2 \rho_{eff}+\rho_r-\delta_r}
\Big]
\end{eqnarray}

By letting $s_2=r_2$ and rearranging the terms, we obtain the following equation to ultimately solve for the stressful peiord $t_s$:
\begin{eqnarray}
s_1 (1+\tilde{\alpha}) &=& (r_1 +\tilde{\alpha} s_1) e^{(\rho_{eff}+\rho_r-\delta_r)t_s} \\
&+& \tilde{\epsilon} \rho_r s_1 
\Big[
\big( \frac{r_1+\tilde{\alpha} s_1}{\rho_{eff}} \big) 
e^{(\rho_{eff}+\rho_r-\delta_r)t_s}
(e^{-\rho_{eff}t_s}-1)
\\ \nonumber 
&+& \frac{\tilde{\alpha} s_1 e^{(\rho_{eff}+\rho_r-\delta_r)t_s}}{2 \rho_{eff}+\rho_r-\delta_r}(1-e^{-(2\rho_{eff}+\rho_r -\delta_r) t_s})
\Big]
\end{eqnarray} 
Assuming $\rho_{eff} t_s \gg 1$, we can approximate the above equation:
\begin{equation}
s_1 (1+\tilde{\alpha}) \simeq [ r_1 +\tilde{\alpha} s_1 -\tilde{\epsilon} \rho_r s_1 H(r_1) ]e^{(\rho_{eff}+\rho_r-\delta_r)t_s}  
+ O(e^{(\rho_r-\delta_r) t_s})
\end{equation}
where $H(r_1) = \frac{r_1+\tilde{\alpha} s_1}{\rho_{eff}}-\frac{\tilde{\alpha} s_1 }{2 \rho_{eff}+\rho_r-\delta_r} = \frac{(r_1+\tilde{\alpha} s_1)(\rho_{eff}+\rho_r-\delta_r)+r_1 \rho_{eff}}{\rho_{eff}(2 \rho_{eff}+\rho_r-\delta_r)}$ is positive when $\rho_{eff} + \rho_r  > \delta_r$.  
Now, the stressful period is given as 
\begin{equation}
t_s =\frac{1}{\rho_{eff}+\rho_r-\delta_r} ln 
\Big( 
\frac{s_1 (1+\tilde{\alpha})}{r_1 +\tilde{\alpha} s_1 -\tilde{\epsilon} \rho_r s_1 H(r_1)}
\Big)
\end{equation}
By inserting $t_s$ back to the Eq. (25), we obtain the relationship between the normalized tumor sizes between the beginning and the ending of the stressful period, $r_1$ and $r_2$:
\begin{equation}
r_2  = \Big( 
\frac{
(s_1)^{\frac{\rho_r -\delta_r}{\rho_{eff}}}
[r_1 +s_1 (\tilde{\alpha}  -\tilde{\epsilon} \rho_r  H(r_1))]}{1+\tilde{\alpha}}
\Big)^{\frac{\rho_{eff}}{\rho_{eff}+\rho_r-\delta_r}}
\end{equation}

\section{Discussion and Conclusion}

We investigated the effects of therapy-induced phenotypic switching of tumor cells on therapeutic efficacy and proposed two therapy strategies, one leading to perpetually oscillating total tumor size and another leading to an asymptotically diminishing total tumor size. Thus, depending on the various clinical situations, we can adopt one of two therapeutic strategies: the former to contain the growth of tumor population if tumor is incurable, and the latter to eliminate tumor cells if tumor is curable. 

We considered the external therapeutic stress as the primary source of tumor phenotypic switching. In principle, the stress can originate from both external and intrinsic sources and, in the absence of external therapeutic stress, the overcrowded cellular environment and the immune response can drive tumor phenotypic switching, the former as the intraspecific competition over limited resources and the latter as the interspecific competition between tumor cells and immune killer cells.  

We considered only two phenotypes of tumor cells and the reversible transition between those two cellular states. In principle, there could be multiple tumor phenotypes (epithelial, mesenchymal, and cellular states between E and M states) and complex transition patterns among those multiple cellular states. This complex transition patterns could be represented by the directed graph models whose nodes represent the cellular states and whose edges represent the transition from one state to another. We suspect that, for the case of a linear chain model of multiple tumor phenotypes, it takes a longer stress-free time to revert from the tolerant to the sensitive tumor phenotype, thus making harder to satisfy the tumor size reduction criterion. 

The effect of the clinically tolerable tumor size $S_M$ on tumor size reduction. What will happen to the final tumor size if we decrease or increase this value of $S_M$? The final non-normalized tumor size is the product of the fixed point $\hat{r}^*$ of the tumor reduction function $F(\hat{r}_1)$ and the normalization factor $S_M$. This fixed point $\hat{r}^*$ is independent of the system size and only dependent on the model parameters (e.g., $\hat{\alpha}$ and $\hat{\beta}$). Suppose that $S_M$ is set larger than the initial resistant tumor size, but small enough that the normalized initial resistant tumor size $\hat{r}_0=R_0/S_M$ is greater than the fixed point $(\hat{r}^*)^{1-\hat{\beta}}$. Then, the normalized resistant tumor size will continue to decrease until it reaches the fixed point $(\hat{r}^*)^{1-\hat{\beta}}$. However, if $S_M$ is too large so that $\hat{r}_0=R_0/S_M$ becomes smaller than the fixed point $(\hat{r}^*)^{1-\hat{\beta}}$, the normalized resistant tumor size increases until it reaches the same fixed point. Thus, if we intend to decrease the resistant tumor size smaller than its initial value, the value of $S_M$ should be chosen to be closer to the initial resistant tumor size. The downside of this choice is that it takes very short rest and stressful time periods initially; i.e., $t_f=-\frac{1}{\rho}ln(\hat{r}_0)$ gets smaller as $\hat{r}_0$ gets closer to one. All the arguments above is applicable only when the tumor reduction criterion is met. 

Finally, the stochastic fluctuations can drive the finite-sized system of tumor cell population to extinction. This is a rare event and the mean passage time to extinction is exponentially proportional to the system size, which implies that it would never happen on a realistic time scale that the system of a large tumor size would go extinct only due to pure stochastic fluctuations. However, it would be possible to use the external perturbation, namely periodic driving, could accelerate the extinction of tumor population.

\appendix*
\section{Tumor size reduction criterion}
We want to find the tumor size reduction criterion: $r_0 (r_1) > r_2 (r_1)$ for $r_1 \in (r^*,1)$. As discussed in details in the prior sections, both functions of $r_0$ and $r_2$ meet at $r_1=1$: $r_0(r_1=1)=r_2(r_1=1)=1$. Also, at $r_1=0$, $r_0 (r_1=0) =0$ while $r_2 (r_1=0)>0$. Thus, since both functions are monotone increasing, the tumor size reduction criterion is equivalent to the condition that two functions intersect at $0<r^*<1$, which requires $\frac{d r_0}{d r_1} < \frac{d r_2}{d r_1}$ at $r_0=\tilde{r}_1=r_1=1$. 

Firstly, given the following functions of $\tilde{r}_1=(r_0)^{\frac{\rho_s+\beta-\rho_r}{\rho_s}}$ and $r_1 = \tilde{r}_1 \Big[
1+\tilde{\epsilon} \rho_r \big(
\frac{(1+\tilde{\beta})}{\rho_s} (r_0 -1) +\frac{\tilde{\beta}}{\beta-\rho_r}(r_0 -\tilde{r}_1)
\big)
\Big]$, we obtain 
\begin{eqnarray}
\Big[ \frac{d r_1}{d r_0} \Big]_{r_0=1} 
&=& \Big[ \frac{d \tilde{r}_1}{d r_0} \Big]_{r_0=1}+\tilde{\epsilon} \rho_r 
\big( \frac{1+\tilde{\beta}}{\rho_s} +\frac{\tilde{\beta}}{\beta-\rho_r} 
(1- \Big[ \frac{d \tilde{r}_1}{d r_0} \Big]_{r_0=1}) \big)
\nonumber
\\ 
&=& \frac{\rho_s +\beta-\rho_r (1-\tilde{\epsilon})}{\rho_s} 
\end{eqnarray}
By taking the inverse of the above equation, we obtain the first derivative of $r_0$ with respect to $r_1$ at $r_0=r_1=1$:
\begin{equation}
\Big[ \frac{d r_0}{d r_1} \Big]_{r_0=1} 
= \frac{\rho_s}{\rho_s +\beta-\rho_r (1-\tilde{\epsilon})}
\end{equation}

Secondly, given the functions of $r_2= \Big( 
\frac{
(s_1)^{\frac{\rho_r -\delta_r}{\rho_{eff}}}
[r_1 +s_1 (\tilde{\alpha}  -\tilde{\epsilon} \rho_r  H(r_1))]}{1+\tilde{\alpha}}
\Big)^{\frac{\rho_{eff}}{\rho_{eff}+\rho_r-\delta_r}}$ 
and
$s_1 = 1+\tilde{\beta}(1-\tilde{r}_1) 
+\tilde{\epsilon} \rho_r \beta 
\Big[ 
(\frac{1+\tilde{\beta}}{\rho_s}+\frac{\tilde{\beta}}{\beta-\rho_r}) 
\frac{r_0 (1-\tilde{r}_1)}{\beta+\rho_s-\rho_r}
+ \frac{(1+\tilde{\beta}) (\tilde{r}_1 -r_0) }{\rho_s(\beta-\rho_r)} 
+ \frac{\tilde{\beta} (\tilde{r}^2_1 -r_0)}{(\beta-\rho_r)(\rho_s+2\beta-2\rho_r)} 
\Big] $ and $H(r_1) = \frac{r_1}{\rho_{eff}} +\frac{\alpha s_1}{\rho_{eff}(2 \rho_{eff}+\rho_r-\delta_r)}$,
we can compute the first derivative of $r_2$ with respect to $r_1$ at $r_1=1$:
\begin{equation}
\Big[ \frac{d r_2}{d r_1} \Big]_{r_0=\tilde{r}_1=1} = g \frac{[1+\tilde{\alpha} -\tilde{\epsilon} \rho_r H(1)]^{g-1}}{(1+\tilde{\alpha})^g}
\Big[
\frac{d(r_1 s^f_1)}{dr_1} +\tilde{\alpha} \frac{d (s^{f+1}_1)}{d r_1} 
-\tilde{\epsilon} \rho_r \frac{d(s^{f+1}_1 H(r_1))}{d r_1}
\Big]_{r_0=\tilde{r}_1=1}
\end{equation}
where $g=\frac{\rho_{eff}}{\rho_{eff}+\rho_r-\delta_r}$, $f=\frac{\rho_r -\delta_r}{\rho_{eff}}$, $s_1(r_0=\tilde{r}_1=1)=1$ and $H(1)=\frac{1}{\rho_{eff}} + \frac{\alpha}{\rho_{eff}(2 \rho_{eff} + \rho_r-\delta_r)}$. The first term inside the bracket on the right hand-side of the above equation is computed as follows:  
\begin{eqnarray}
\Big[ \frac{d(r_1 s^f_1)}{dr_1} \Big]_{r_0=\tilde{r}_1=1} 
&=& \Big[s^f_1 \Big]_{r_0=\tilde{r}_1=1}  + \Big[ r_1 f s^{f-1}_1 
\frac{d s_1}{d r_0} \frac{d r_0}{d r_1} \Big]_{r_0=\tilde{r}_1=0} 
\\ \nonumber
&=& 1- \big( \frac{\rho_r-\delta_r}{\rho_{eff}} \big) \big( \frac{\beta}{\rho_s +\beta-\rho_r (1-\tilde{\epsilon})} \big)
\end{eqnarray}
where 
\begin{eqnarray}
\Big[ \frac{d s_1}{d r_0} \Big]_{r_0=\tilde{r}_1=1} &=& 
-\tilde{\beta} \Big[ \frac{d \tilde{r}_1}{d r_0} \Big]_{r_0=1} 
+\tilde{\epsilon} \rho_r \beta 
\Big[ 
(\frac{1+\tilde{\beta}}{\rho_s}+\frac{\tilde{\beta}}{\beta-\rho_r}) 
(\frac{1-\tilde{r}_1 - r_0 \frac{d \tilde{r}_1}{d r_0}}{\beta+\rho_s-\rho_r})
\\ \nonumber
&+& \frac{(1+\tilde{\beta}) (\frac{d \tilde{r}_1}{d r_0} - 1) }{\rho_s(\beta-\rho_r)} 
+ \frac{\tilde{\beta} (2 \tilde{r}_1 \frac{d \tilde{r}_1}{d r_0} - 1)}{(\beta-\rho_r)(\rho_s+2\beta-2\rho_r)} 
\Big]_{r_0=\tilde{r}_1=1}
\\
&=&
-\tilde{\beta} (\frac{\rho_s +\beta -\rho_r}{\rho_s})
+\tilde{\epsilon} \rho_r \beta 
\Big[-\frac{1}{\rho_s} (\frac{1+\tilde{\beta}}{\rho_s} +\frac{\tilde{\beta}}{\beta-\rho_r}) +\frac{1+\tilde{\beta}}{\rho^2_s} +\frac{\tilde{\beta}}{\rho_s (\beta-\rho_r)} 
\Big]
\\ \nonumber
&=& -\frac{\beta}{\rho_s}
\end{eqnarray}
The second term inside the bracket on the right hand-side of the above equation is computed as follows:  
\begin{equation}
\tilde{\alpha} \Big[ \frac{d (s^{f+1}_1)}{dr_1} \Big]_{r_0=\tilde{r}_1=1} 
= \tilde{\alpha} (f+1) \Big[ s^f_1 \frac{d s_1}{d r_0} \frac{d r_0}{d r_1}\Big]_{r_0=\tilde{r}_1=1} 
= -\frac{\alpha \beta}{\rho_{eff} (\rho_s+\beta-\rho_r (1-\tilde{\epsilon}))}
\end{equation}
The third term inside the bracket on the right hand side is computed as follows:
\begin{eqnarray}
\Big[ \frac{ d (s^{f+1}_1 H(r_1))}{d r_1} \Big]_{r_0=\tilde{r}_1=1} 
&=& (f+1) \big[ s^f_1 H(r_1) \frac{d s_1}{d r_0} \frac{d r_0}{d r_1}\big]_{r_0=r_1=1} + \Big[ s^{f+1}_1  \frac{d H(r_1)}{d r_1} \Big]_{r_0=r_1=1}
\\ \nonumber
&=& -(\frac{\rho_{eff} +\rho_r -\delta_r}{\rho_{eff}}) (\frac{1}{\rho_{eff}} + \frac{\alpha}{\rho_{eff}(2 \rho_{eff} + \rho_r-\delta_r)})(\frac{\beta}{\rho_s +\beta -\rho_r (1-\tilde{\epsilon})})
\\ \nonumber
&+& \frac{1}{\rho_{eff}} \Big[ 1-
\frac{\alpha \beta}{ (2 \rho_{eff} +\rho_r -\delta_r)(\rho_s +\beta -\rho_r (1-\tilde{\epsilon}))} \Big]
\end{eqnarray}
where 
\begin{eqnarray}
\Big[ \frac{d H(r_1)}{d r_1} \Big]_{r_0=r_1=1} 
&=& \frac{1}{\rho_{eff}} +
\frac{\alpha}{\rho_{eff}(2 \rho_{eff}+\rho_r-\delta_r)} \frac{d s_1}{d r_0} \frac{d r_0}{d r_1} 
\\ \nonumber
&=& \frac{1}{\rho_{eff}} -
\frac{\alpha \beta}{\rho_{eff}(2 \rho_{eff}+\rho_r-\delta_r)(\rho_s +\beta-\rho_r (1-\tilde{\epsilon}))}
\end{eqnarray}

Now, the first derivative of $r_2$ with respect to $r_1$ at $r_1=1$ is given as follows:
\begin{eqnarray}
\Big[ \frac{d r_2}{d r_1} \Big]_{r_0=\tilde{r}_1=1} 
&=& 
g \frac{[1+\tilde{\alpha} -\tilde{\epsilon} \rho_r H(1)]^{g-1}}{(1+\tilde{\alpha})^g}
\Big\{
1- \frac{\beta (\alpha +\rho_r -\delta_r)}{\rho_{eff}(\rho_s +\beta-\rho_r (1-\tilde{\epsilon}))}
\Big[
1-\tilde{\epsilon} \rho_r Q 
\Big]
\Big\}
\end{eqnarray}
where 
$Q=\frac{1}{\alpha +\rho_r -\delta_r}(\frac{\rho_r-\delta_r}{\rho_{eff}} 
+ \frac{2 \alpha}{2 \rho_{eff} +\rho_r -\delta_r (1-\tilde{\epsilon})} 
+\frac{\alpha (\rho_r-\delta_r)}{\rho_{eff}(2 \rho_{eff} +\rho_r -\delta_r)}
-\frac{\rho_s -\rho_r (1-\tilde{\epsilon})}{\beta}
)$.

\begin{figure}[htbp]
\begin{center}
\includegraphics[width=1\textwidth]{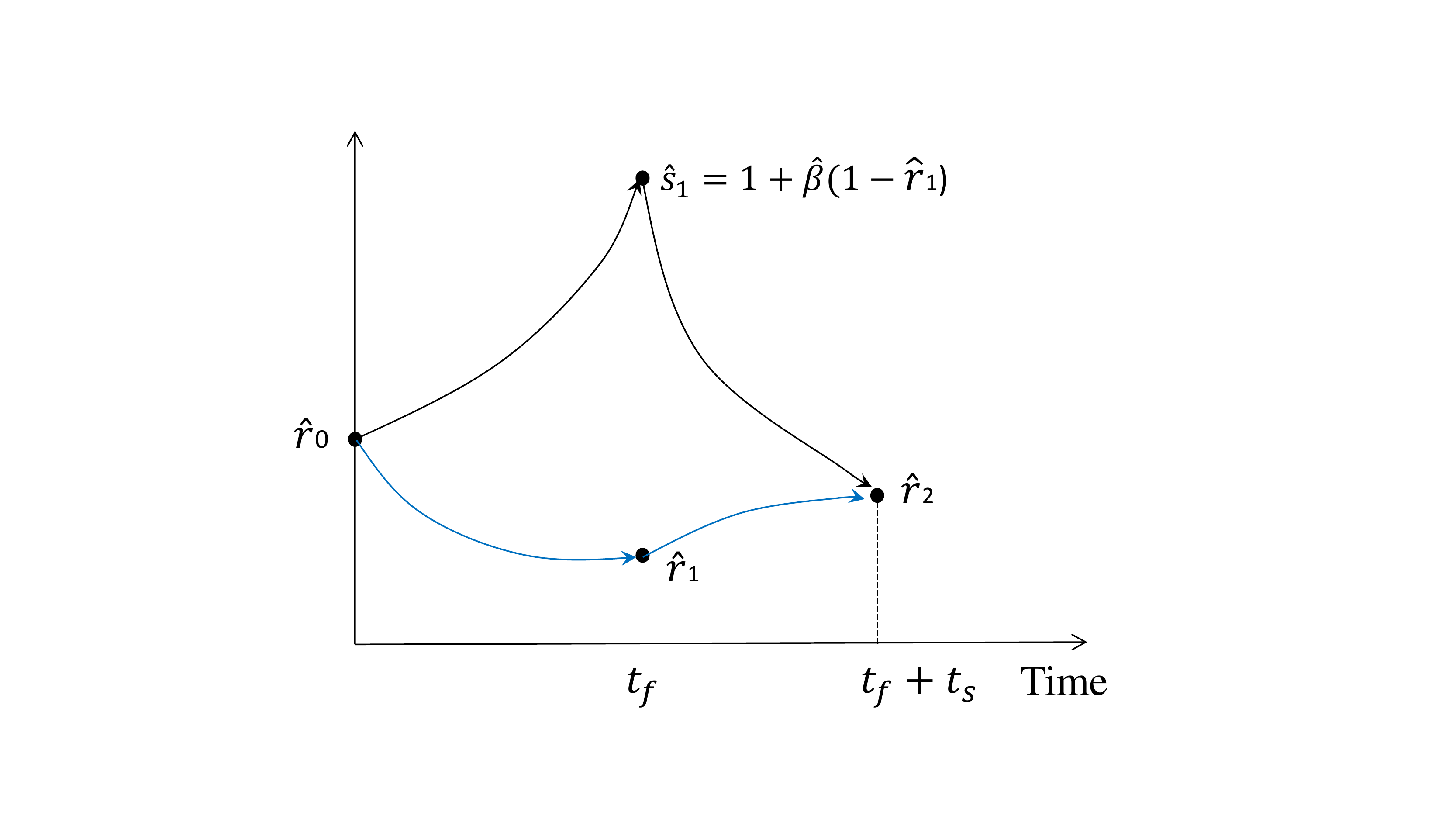} 
\caption{The dynamics of two tumor phenotype sizes in a single cycle of therapy. Black (red) curve represents the time evolution of the sensitive (resistant) tumor phenotype size. Both tumor phenotype sizes are normalized by a constant and fixed value of a clinically tolerable tumor load size $S_M$. Both tumor sizes are equal at the beginning (denoted by $\hat{r}_0$) and the ending (denoted by $\hat{r}_2$) of a single therapeutic cycle. The cycle starts with a stree-free environment which lasts for the duration of $t_f$ and ends with a stressful/therapuetic environment for the period of $t_s$.} 
\label{Fig1}
\end{center}
\end{figure}

\begin{figure}
\begin{center}
\includegraphics[width=1\textwidth]{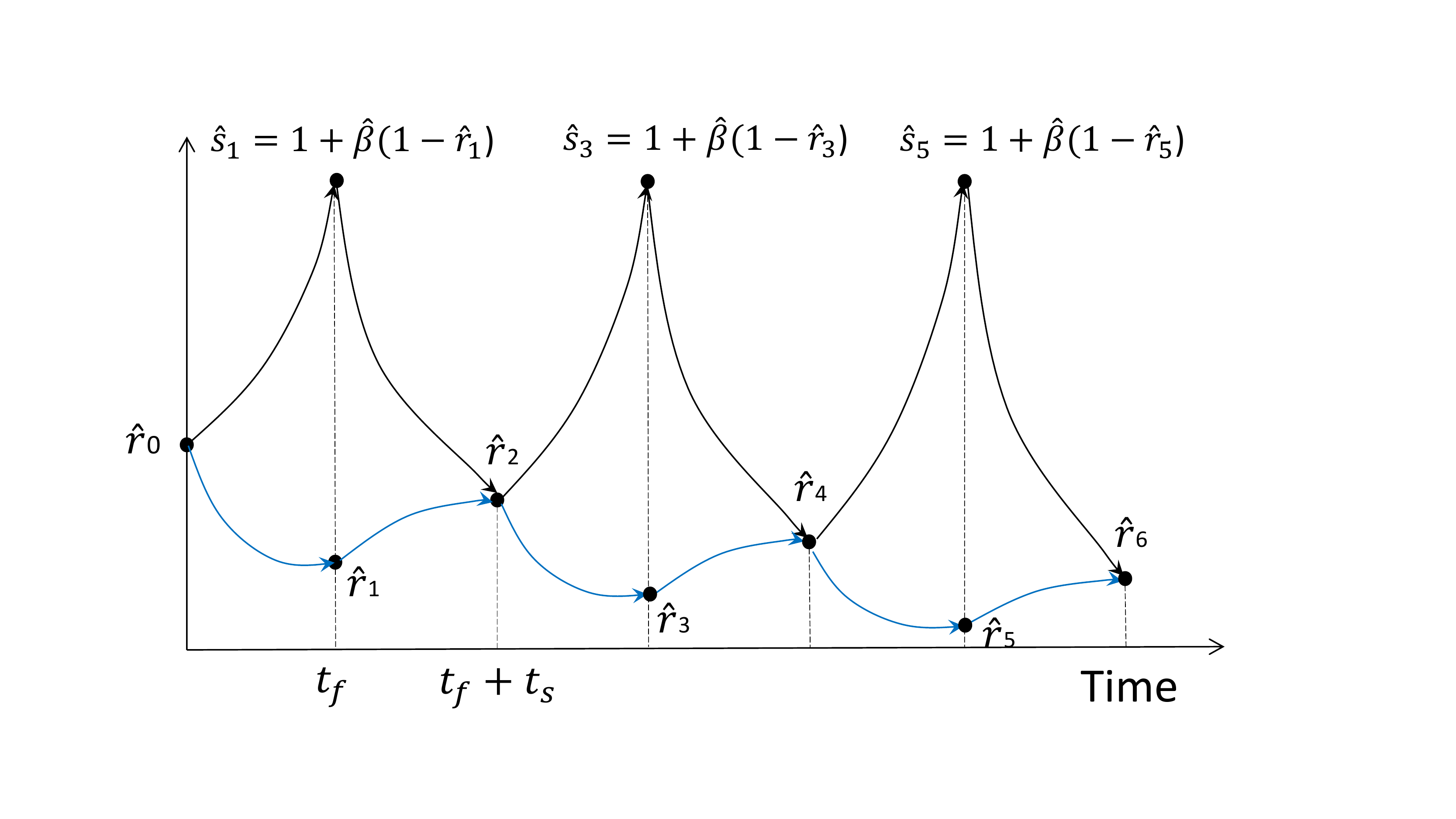} 
\caption{Time evolution of the normalized tumor phenotype sizes during multiple cycles of therapy. Black (blue) curve represents the time evolution of the sensitive (resistant) tumor phenotype size. At the beginning and the ending of each of $(j+1)$th cycle, both sensitive and resistant phenotype tumor sizes coincide, e.g., $\hat{r}_{2j}=\hat{s}_{2j}$ for $j=0,1,...N$. During each stress-free interval of $(j+1)$th cycle, the normalized sensitive tumor phenotype is allowed to reach the clinically tolerable tumor load of $1+\hat{\beta}(1-\hat{r}_{2j+1})$.}
\label{Fig2}
\end{center}
\end{figure}

\begin{figure}
\begin{center}
\includegraphics[width=1\textwidth]{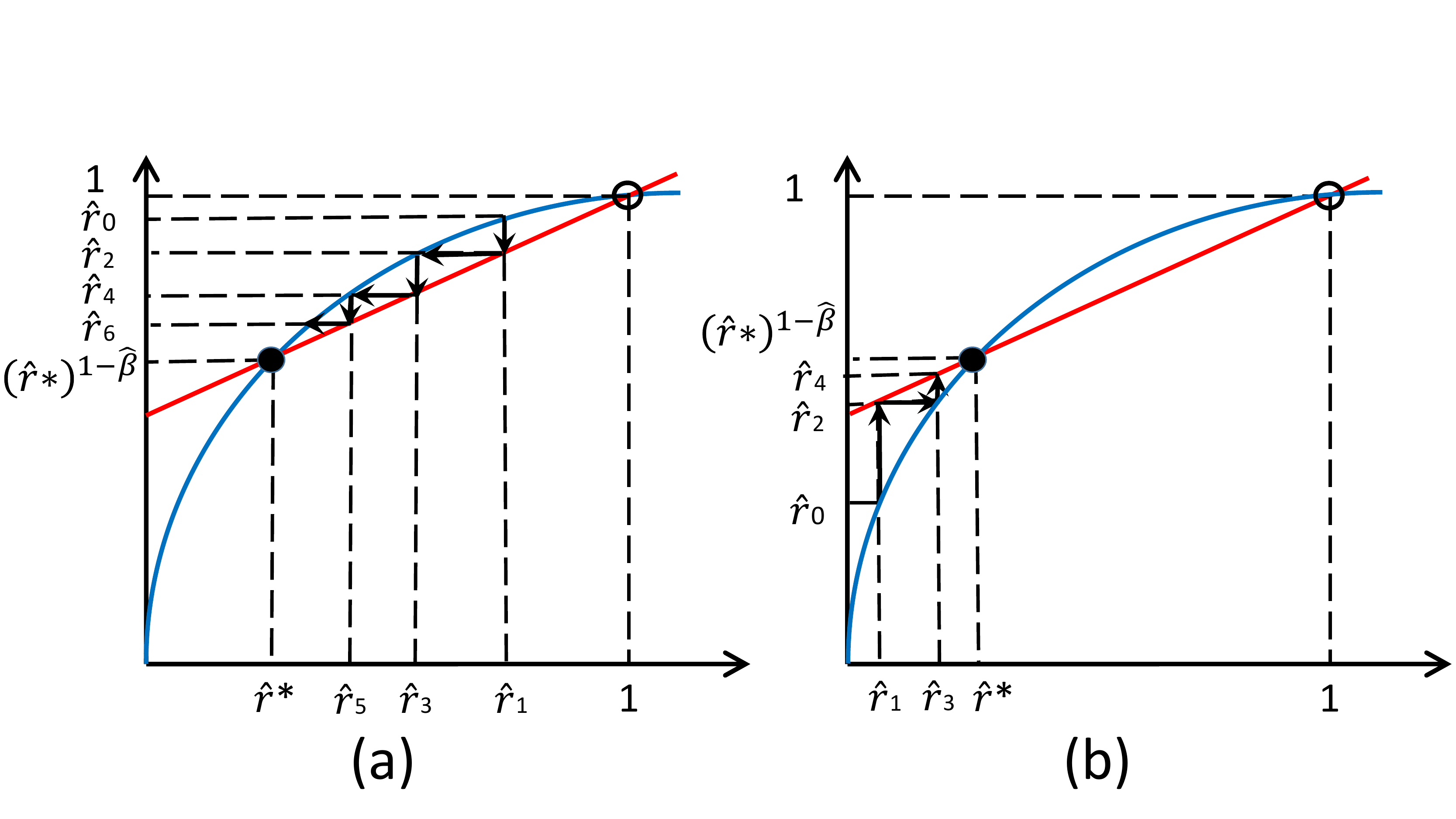} 
\caption{Iterated map of the progression of the normalized resistant tumor sizes after multiple cycles of therapy when the tumor reduction criteria are satisfied. (a) When the initial normalized resistant tumor size of $\hat{r}_0$ is greater than $\hat{r}^*$. (b) When $\hat{r}_0<\hat{r}^*$. The normalized resistant tumor sizes after each cycle of therapy are determined based on the updating rules in Eq. (16) and (17). The blue line represents the resistant tumor size updating rule for $\hat{r}_{2j} \rightarrow  \hat{r}_{2j+1}$ and the red line depicts the resistant tumor size updating rule for $\hat{r}_{2j+1} \rightarrow \hat{r}_{2j+2}$. Given the value of the initial normalized resistant tumor size of $\hat{r}_0$, one draws a horizontal line touching $\hat{r}_0$ and determines the value of $\hat{r}_1$ from the intersection point between the blue line and the horizontal line touching $\hat{r}_0$. Then, one draws a vertical line touching $\hat{r}_1$ and determines the value of $\hat{r}_2$ from the interesection point between the red line and the vertical line touching $\hat{r}_1$. We repeat the forementioned mapping process over and over again for each subsequent cycle of therapy. Such an iterated mapping let $\hat{r}_{2N}$ converges to $(\hat{r}^*)^{1-\hat{\beta}}$ in the limit of large number of cycles, regardless of the initial tumor size.}
\label{Fig3}
\end{center}
\end{figure}

\begin{figure}
\begin{center}
\includegraphics[width=1\textwidth]{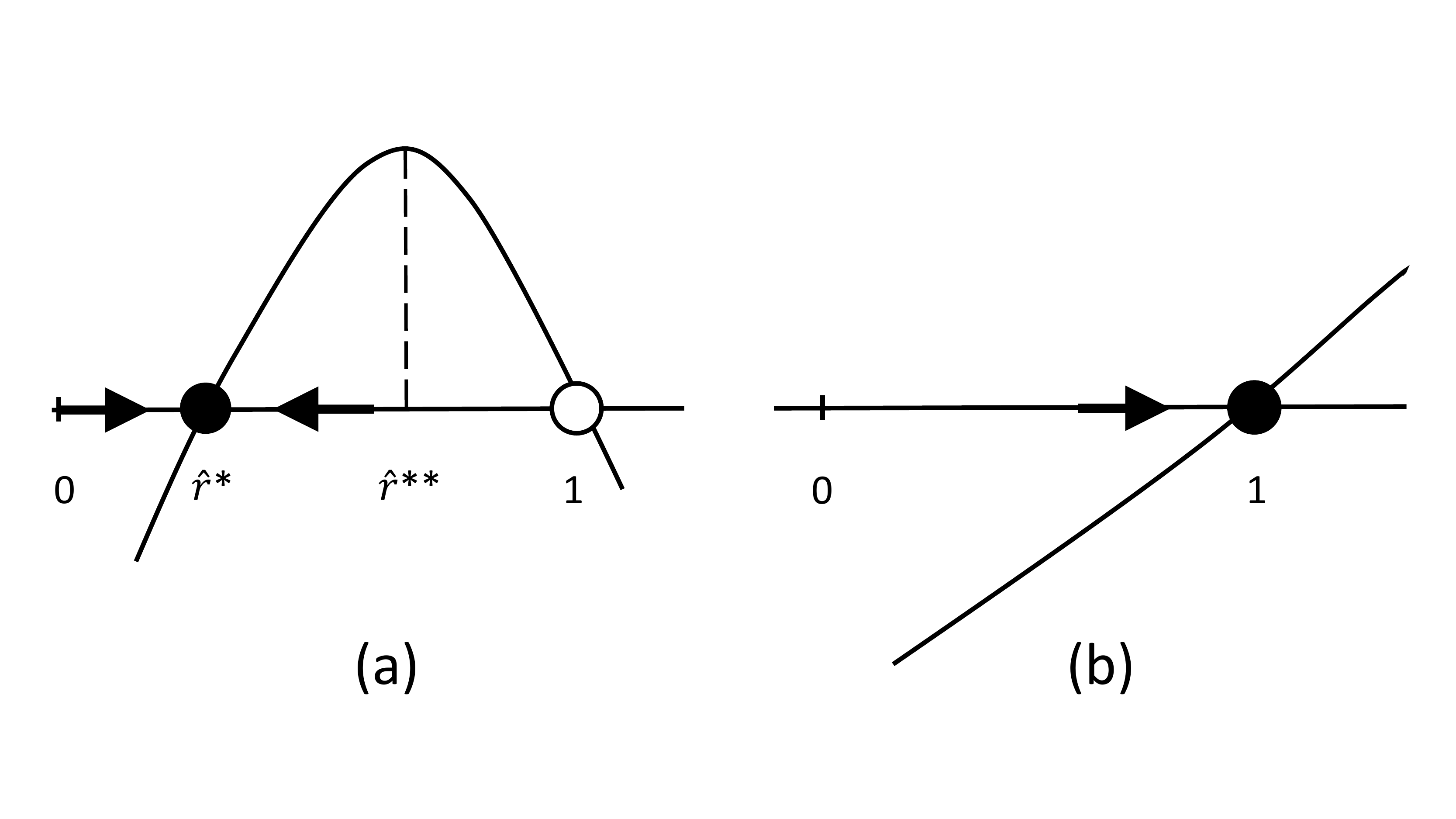} 
\caption{Fixed points of the tumor reduction function $F(\hat{r}_1)$. (a) When $\hat{\alpha}< \hat{\beta}$, there are two fixed points, a stable fixed point at $\hat{r}_1=\hat{r}^*$ and an unstable fixed point at $\hat{r}_1=1$. The tumor reduction function $F(\hat{r}_1)$ is maximal at $\hat{r}_1=\hat{r}^{**}$. (b) When $\hat{\alpha}> \hat{\beta}$, there is one stable fixed point at $\hat{r}_1=1$. Open (filled) dot indicates the unstable (stable) fixed point.}
\label{Fig4}
\end{center}
\end{figure}

\begin{figure}
\begin{center}
\includegraphics[width=1\textwidth]{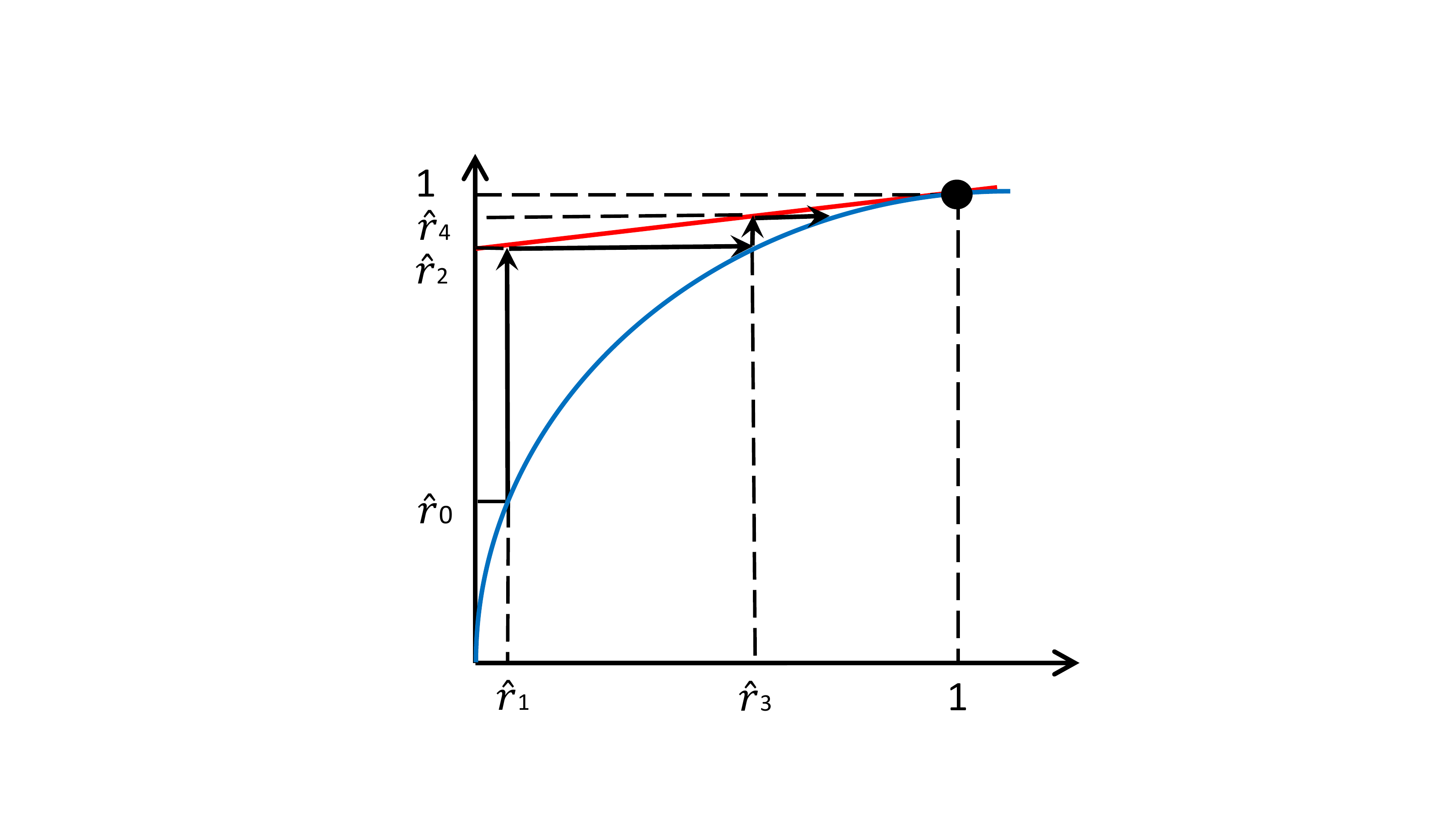} 
\caption{Iterated map of the progression of the normalized resistant tumor sizes after multiple cycles of therapy when the tumor reduction criteria are violated. The normalized resistant tumor sizes are updated based upon the updating rules in Eq. (16) and (17).}
\label{Fig5}
\end{center}
\end{figure}

\begin{figure}
\begin{center}
\includegraphics[width=1\textwidth]{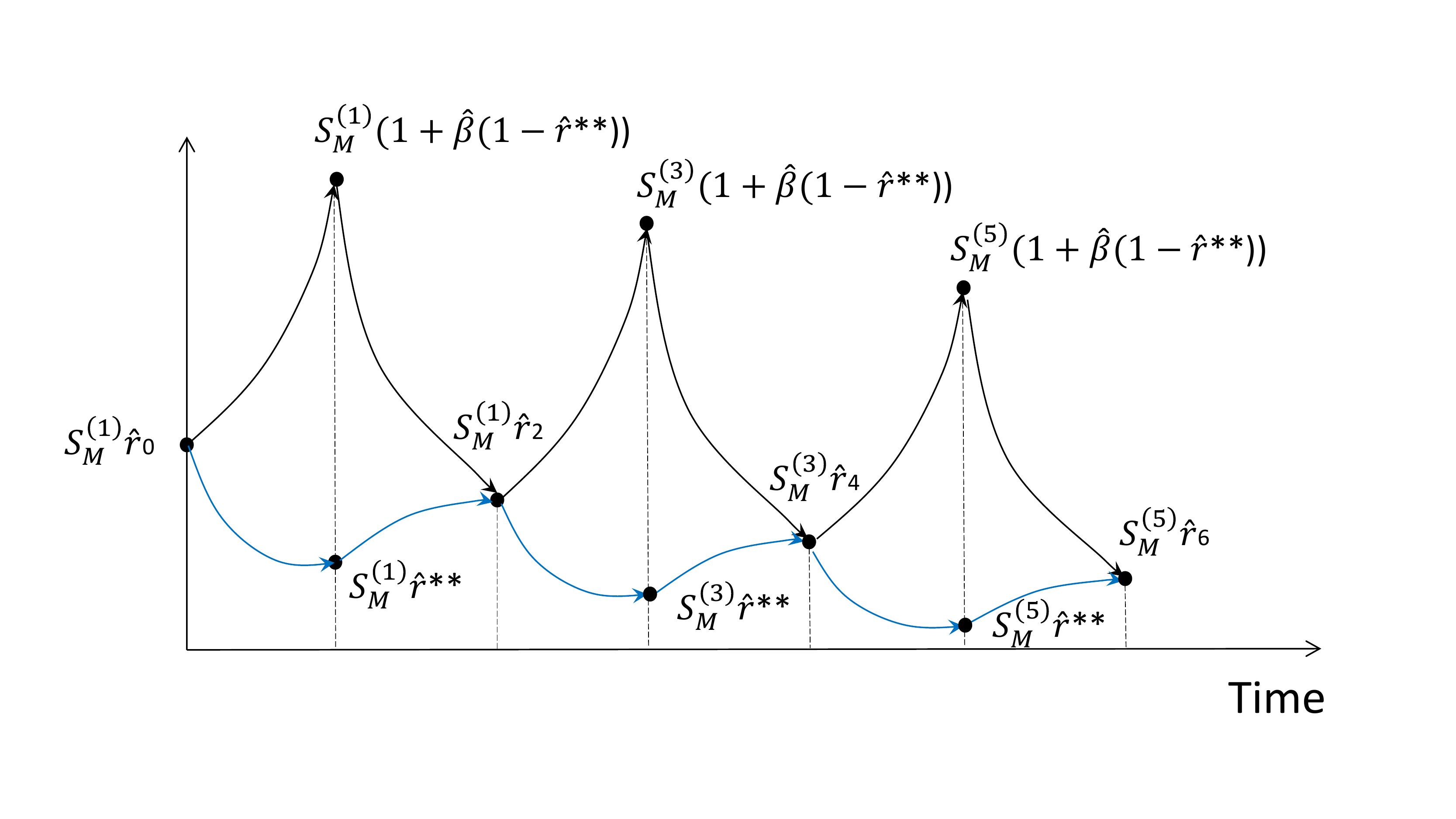} 
\caption{Time evolution of the normalized tumor phenotype sizes during multiple cycles of therapy under the therapy strategy II with the maximal tumor size reduction condition. Black (blue) curve represents the time evolution of the sensitive (resistant) tumor phenotype sizes. The normalized resistant tumor phenotype size is determined to reach $\hat{r}^{**}$ at the end of each stress-free environment. During each stress-free interval, the sensitive tumor phenotype size is allowed to reach the variable tumor load of $S^{2j+1}_M$ and both tumor phenotype sizes are normalized by the normalization factor of $S^{2j+1}_M$ for the $(j+1)$th cycle $j=0,1,2,...N-1$.}
\label{Fig6}
\end{center}
\end{figure}

\begin{figure}
\begin{center}
\includegraphics[width=1\textwidth]{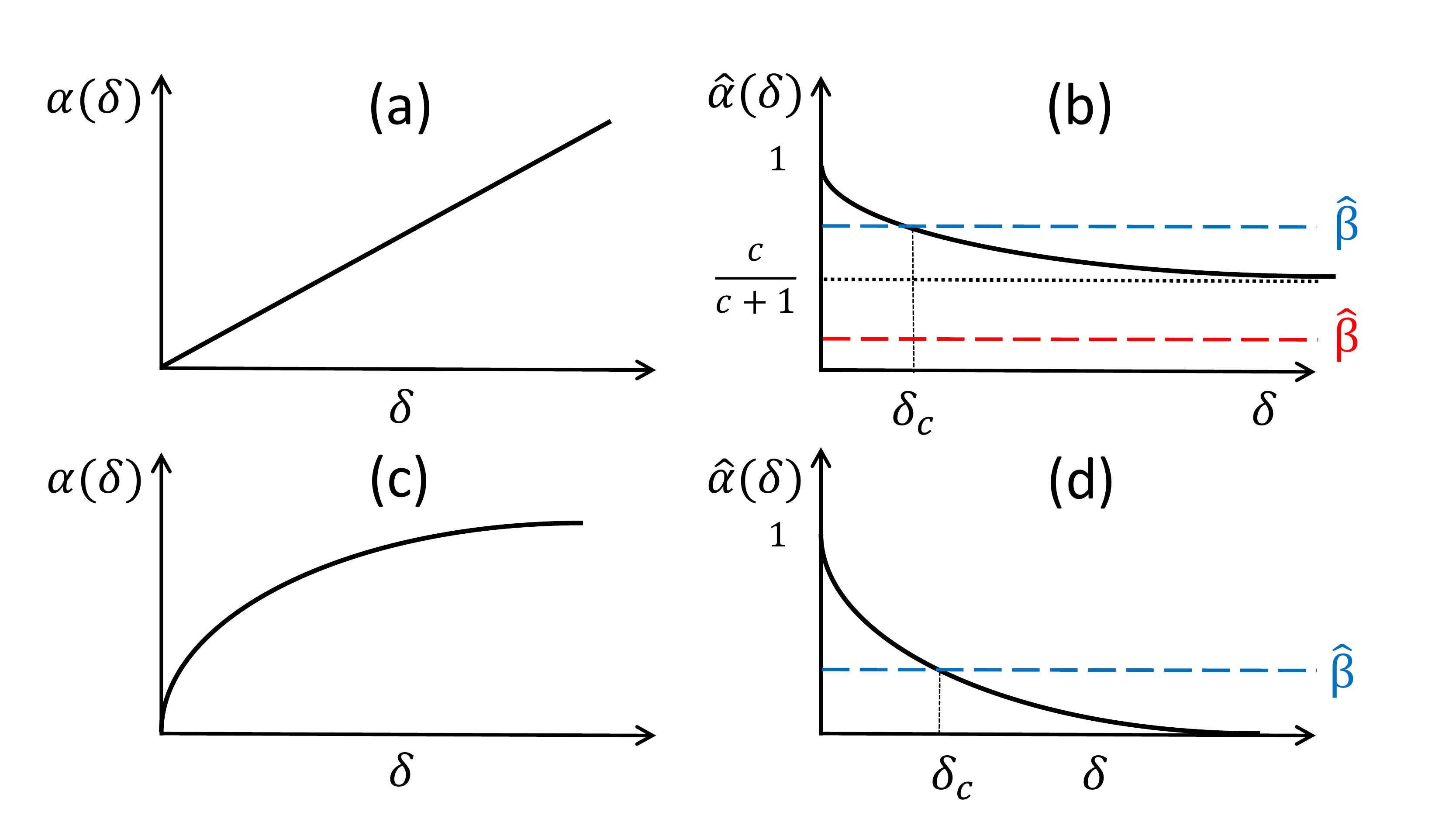} 
\caption{Effects of stress level-dependent forward switching rate on tumor size reduction criterion. (a) Linear case: $\alpha(\delta)=c\delta \Theta(\delta-\rho)$, (b) $\hat{\alpha}(\delta)$ is a monotonic decreasing function of $\delta$ and saturates to a constant $\frac{c}{c+1}$ in the large $\delta$ limit. 
For the blue colored dashed line of $\hat{\beta}$ ($\hat{\beta} > \frac{c}{c+1}$), the tumor reduction criterion, $\hat{\beta} > \hat{\alpha}$, is satisfied when $\delta>\delta_c$. For the red-colored dashed line  of $\hat{\beta}$ ($\hat{\beta} < \frac{c}{c+1}$), the tumor reduction criterion, $\hat{\beta} > \hat{\alpha}$, cannot be satisfied for any value of $\delta$. (c) Saturating case: $\alpha=\frac{c \delta \Theta(\delta-\rho)}{\delta+d}$, (d) $\hat{\alpha}(\delta)$ is a monotonic decreasing function of $\delta$ and approaches zero in the large $\delta$ limit. Regardless of the value of $\hat{\beta}$, the tumor size reduction criterion of $\hat{\alpha}<\hat{\beta}$ is satisfied for $\delta > \delta_c$.  
}
\label{Fig7}
\end{center}
\end{figure}

\begin{figure}
\begin{center}
\includegraphics[width=1\textwidth]{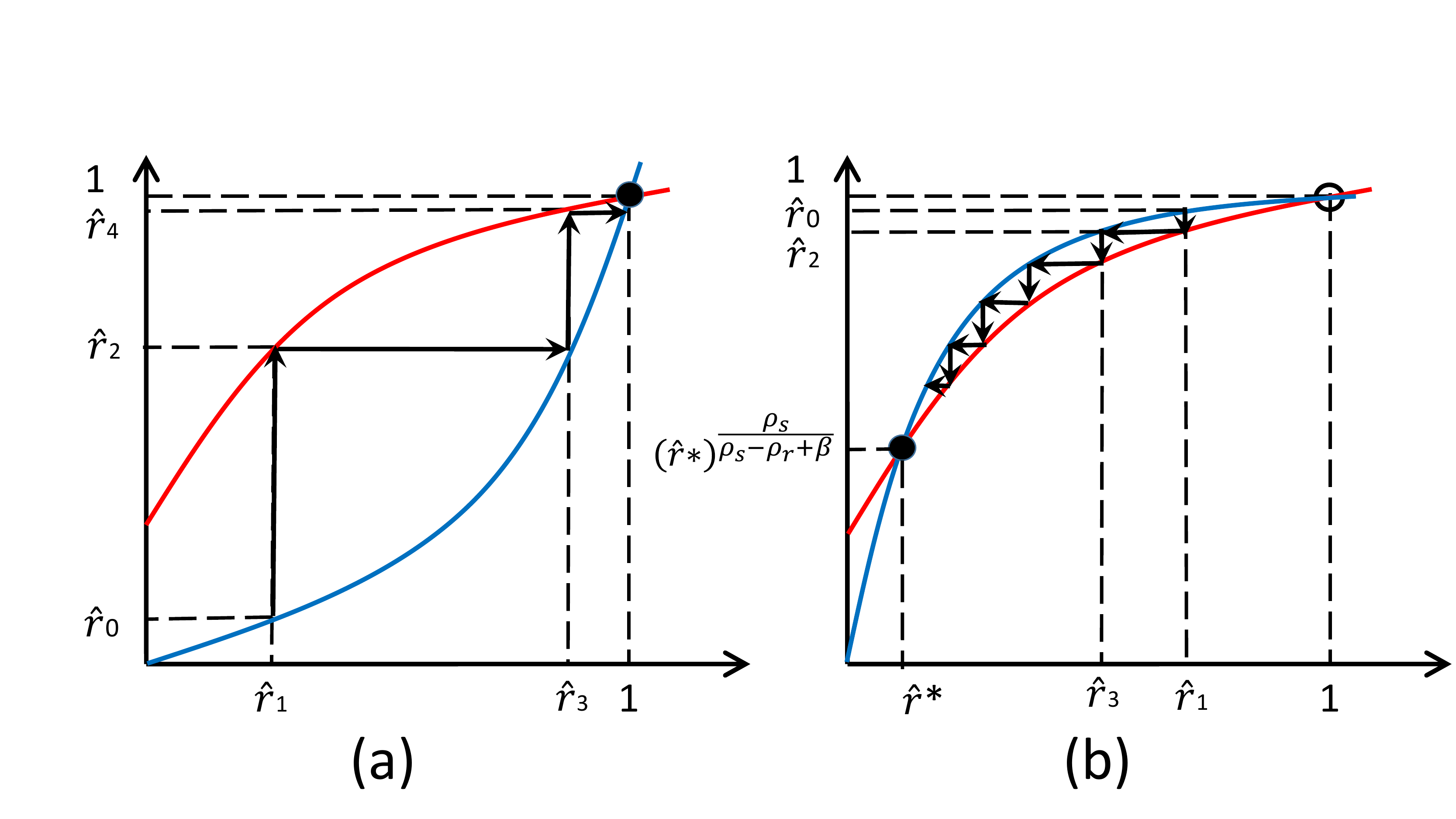} 
\caption{Iterated map of the progression of normalized resistant tumor sizes after multiple cycles of therapy for the case of non-zero growth rate of resistant phenotype. The blue line represents the resistant tumor size updating rule for $\hat{r}_{2j} \rightarrow  \hat{r}_{2j+1}$ and the red line depicts the resistant tumor size updating rule for $\hat{r}_{2j+1} \rightarrow \hat{r}_{2j+2}$ where $j=0,1,...,N-1$.  Filled (open) circles represent the stable (unstable) fixed point of the iterated map. (a) When $\beta < \rho_r$, the normalized resistant tumor sizes approach closer to $1$ after each cycle of therapy. (b) When $\beta > \rho_r$, the normalized tumor sizes move away from 1 (an unstable fixed point) and approach closer to the second intersection point (a stable fixed point) between blue and red lines after each cycle of therapy. 
}
\label{Fig8}
\end{center}
\end{figure}

\end{document}